# Six decades of TIGRE and Mount Wilson chromospheric monitoring in the H and K lines: The quest for an understanding of solar-type activity

J. H. M. M. Schmitt[1], K.-P. Schröder[2], M. Mittag[1], A. Hempelmann[1], J. N. González-Pérez[1], Dennis Jack[2]

1: Hamburger Sternwarte, Universität Hamburg, Gojenbergsweg 112, 21029 Hamburg, Germany
2: Universidad de Guanajuato, Callejon de Jalisco S/N, 36023 Guanajuato, GTO, Mexico

**ABSTRACT**

The combined time series of the consistently calibrated chromospheric activity indicator, the S-index – derived from Mount Wilson and TIGRE data – now spans more than five decades of monitoring observations of the Sun and of more than one hundred solar-like stars. For the first time, these data allow us to probe the long-term stability of solar-type magnetic cycles as well as their possible transition into other behavioral forms of stellar activity in the time domain. Such variability studies are an important input for solar-stellar connection research and for our empirical understanding of the long-term perspective of solar activity as well as of a wider empirical picture of the coevolution of magnetic activity with stellar structure. We here present and discuss a representative selection of cool main-sequence stars, for which such data are available. The excellent quality of our TIGRE data now more than doubles the number of known cyclic stars. Grading them (i.e., excellent, good, fair, and poor) and putting them into an evolutionary perspective, we find that stars more active than the Sun are also more likely to show cyclic variability than stars below solar activity. Furthermore, more active stars tend to have more regular cycles, while the solar mean S-value fits the lower end of the S-index range found for cyclic stars. This is consistent with the occurrence of solar Maunder minima episodes, hinting at a sometimes already unstable solar cycle. We find eighteen stars with even lower activity and lower variability than the Sun, which provide a preview into the latter's distant future. For half of these stars, we can use sensitive X-ray observations to complement our chromospheric data and to explore the lower end of stellar activity also from a coronal perspective. These so-called "Maunder minimum candidate stars" exhibit what we refer to as "remnant activity," which expresses itself in a minimal coronal flux and a minimal S-value, above which only rather small variations occur. Using accurate *Gaia* DR3 distances allows us to place our sample stars into an Hertzsprung-Russell diagram (HRD), which demonstrates that active cyclic stars and inactive noncyclic stars populate rather different regions in the HRD.

## 1. Introduction

The cyclic appearance of Sun spots on the solar surface was first clearly described by Schwabe (1844), yet at that time their magnetic character was unknown. Today's common understanding is that Sun spots are just one manifestation of what is generally referred to as "magnetic activity." On the Sun such "magnetic activity" can be observed over the whole electromagnetic spectrum, ranging from $\gamma$ rays to radio emission. All ubiquitously exhibit, on average, an 11-year periodicity in the recorded signals from the effects of magnetic activity in all layers of the atmosphere of the Sun, i.e., the photosphere, chromosphere, and corona. One may ask to what extent such cyclic "magnetic activity" is also present on other stars. In contrast to the Sun, stars can (with very few exceptions) not be spatially resolved, and therefore any "magnetic activity" phenomena must be detected against the background of the full but unresolved stellar disk. Broadband photometric time series very often remain ambiguous, because their variations are typically rather small and react not only to star spots but also to faculae and, possibly, to pulsations.

A very clear signature of the activity level of a given star is the strength of the chromospheric emission in the cores of the Ca II H and K lines. Such line reversals were first noted to be associated with active regions over a hundred years ago by Eberhard & Schwarzschild (1913), who reported line reversals in the cores of the Ca II H and K absorption lines of Arcturus, Aldebaran, $\sigma$ Gem, and other stars. They also noted the similarity of these features to those observed in active solar regions. Eberhard & Schwarzschild (1913) even conjectured about possible intensity variations in these line reversals, analogous to the solar cycle, which could be explored via systematic monitoring.

However, this idea was not followed up until half a century later when O.C. Wilson started his chromospheric activity monitoring project in the 1960s, introducing a simple but well calibrated "S-index" (see Sect. 3.1), which measures the strength of the Ca II H and K line core emission relative to the adjacent continuum with just one number. By its construction it is basically independent of any sky transparency issues. Using this S-index as a measure of the fluxes in the centers of the Ca II H and K lines, Wilson (1978) presented the first long-term chromospheric time series (extending roughly over a decade) of a larger sample of stars (91 objects at first), studied systematic variations, and identified one star (i.e., HD 81809) with a definite solar-like Ca II H and K cycle. Then, in a landmark paper, Baliunas et al. (1995) presented Mount Wilson S-index time series observations of 111 typically main-sequence stars of spectral type F2 - M2, with the bulk of stars in the spectral ranges, F and G. The time span covered by these observations is usually about 25 years from the mid-1960s to the mid-1990s. For a subsample of 35 stars, data are available until 2001, and some of these data are presented by Oláh et al. (2016).

The Mount Wilson data, forming the backbone of our work, were essentially "hand collected." Our TIGRE work presented here is intended as an extension in time of the Mount Wilson chromospheric activity monitoring project. However, we use modern approaches utilizing CCD detectors, fully robotic operations, and internet technology. These allow us to achieve higher observing cadences and precision, which in turn allow us to probe shorter cycle lengths, cycle (in)stability, and a better characterization of smaller, noncyclic activity-variation of less active stars over time.

Thus, in this paper we examine the chromospheric properties of these stars over these longer timescales, i.e., more than five cycles in terms of the solar cycle length. In addition, we can now use additional information not available in the context of the Mount Wilson program (i.e., data obtained at other wavelength ranges most notably in the X-ray band) and, most importantly, the now available very precise *Gaia* DR3 parallaxes, which enable – together with the also very much improved stellar metallicities – an accurate placement of our sample stars in the Hertzsprung-Russell diagram (HRD) as well as meaningful comparisons both with model calculations and the Sun. This wealth of new data enables us to develop a picture of the coevolution of magnetic activity with the star and how it depends on stellar mass and/or the thickness of the convective envelope.

The specific plan of our paper is as follows. In Sect. 2 we provide an overview of stellar magnetic activity diagnostics over different spectral bands. In Sect. 3 we present the underlying observations, the stellar sample used, and our data analysis procedures. Section 4 is devoted to the long-term S-index time series of the Sun. In Sect. 5 we describe the construction of the long-term calibrated stellar S-index time series. In Sect. 6 we discuss the stability of stellar cycles over 60 years, and in Sect. 7 we compare the activity properties of cyclic and noncyclic stars, introducing Maunder minimum candidate stars and discussing their properties from the chromospheric and coronal perspective. In Sect. 9 we place the stars in an HRD and study the respective locations of cyclic and noncyclic stars. We draw our conclusions in Sect. 10. Note that for better readability, we moved a substantial number of figures and tables, as well as supplementary material, to the appendices.

## 2. Stellar magnetic activity diagnostics

Solar magnetic activity can be diagnosed almost over the whole electromagnetic spectrum. The 10.6 cm coronal radio flux, a side-channel of coronal emissions well known from ground-based solar radio observations, has been observed continuously for almost 60 years. Yet, the radio emission of many stars is rather weak, so that only stars in the immediate solar neighborhood can be studied in this wavelength range.

Another very clear signature of magnetic activity is provided by coronal X-ray emission, which is an excellent proxy indicator for "magnetic activity" phenomena in cool stars, since an X-ray "background" of a pristine stellar photosphere is completely absent. Yet, X-ray observations need space-borne instrumentation, which makes long-term monitoring rather expensive and complicated. Also, careful cross-mission calibrations are required. While the first published X-ray surveys of cool stars had sample sizes of a bit over a hundred objects (Vaiana et al. 1981), recent X-ray surveys conducted by eROSITA yield active star samples with sizes of several hundred thousand objects (Freund et al. 2024). Unfortunately, the opportunities for long-term monitoring in X-rays over timescales of tens of years are very much limited, in fact, such long-term X-ray monitoring exists only for very few individual sources (Robrade et al. 2012; Orlando et al. 2017).

Chromospheric activity on the other hand can be straightforwardly studied from the ground, preferably using spectroscopy of chromospheric emission lines such as the Ca II H and K lines, the Ca II infrared triplet (IRT) lines, and the H$\alpha$ lines to name just a few. The most extensive set of chromospheric activity studies uses the S-index introduced by O. Wilson and his collaborators (see Sect. 3.1 for details). Systematic measurements of S-indices and their time evolution were carried out at the Mount Wilson Observatory near Pasadena, referred to as Mount Wilson HK project in the following. Apart from monitoring selected solar-like stars and cool giants, exemplary S-index values were measured for about 3000 stars several times to compare their average activity levels. Those data collected in the framework of the Mount Wilson HK project have been published in summary form by Duncan et al. (1991), who also provide a description of the evolution of the instrumentation used. For a smaller set of 111 stars, published in the already mentioned paper by Baliunas et al. (1995), relatively densely sampled S-index light curves are available. There are other S-index monitoring efforts, most notably those carried out with the Solar Stellar Spectrograph (SSS), which covers only 72 stars with typically lower sampling cadence (Radick et al. 2018). Chromospheric activity in cool stars appears to be very well correlated with coronal activity (for a recent study see Fuhrmeister et al. (2022) and references therein), and we note in passing that HD 81809 also shows a very clear X-ray activity cycle (Favata et al. 2008; Orlando et al. 2017) and an extremely tight correlation between its chromospheric and coronal emissions.

O.C. Wilson recognized the importance of Sun-as-a-star observations for providing a well-understood comparison star to stellar observations. By using the moon as a reflector for solar light, O.C. Wilson could observe the solar chromospheric activity level using the same instrumentation as for his sample of over a hundred solar-like stars and about 40 hardly variable stars that served as calibrators. This is key to making S-index monitoring independent of instrumentation, enabling totally different types of equipment to continue the time series and produce cross-calibrated results that can be meaningfully compared.

Yet, the S-index as introduced by O.C. Wilson and his collaborators is a purely instrumental index, and furthermore, it contains contributions from the photosphere, which vary, for example, with the effective temperature of the underlying star, and are not related to stellar activity. As a result, it is not straightforward to quantitatively compare the activity of different stars based on their S-indices and extract physical information in the form of energy fluxes and their variations from these data. The extraction of such physical quantities from S-index data became only possible, when Linsky et al. (1979), Middelkoop (1982), and Rutten (1984) demonstrated how to convert the instrumental S-indices into physical surface fluxes, which enabled comparisons between stars of different spectral types and an assessment of the energetics of the observed chromospheric Ca II H and K emission.

The whole database collected in the context of the Mount Wilson HK project has been made public and available to the astronomical community at large [1], and a detailed description of this HK dataset has been provided by R. Radick [2]. For a smaller subset of 35 stars additional data covering the period until 2001 is publicly available and can be also downloaded [3].

The data released by NSO are not fully presented in the paper by Baliunas et al. (1995), which covers only the period between 1966 and 1992. As described by Vaughan et al. (1978), the Mount Wilson HK observations were carried out "by hand," and the observing cadence is typically lower during the early years and higher during the 1980s; for a more detailed description of the sample and the observations carried out with the Mount Wilson H and K photometers we refer the reader to Baliunas et al. (1995) and the references therein. The fundamental results of the study by Baliunas et al. (1995) are provided in Table 2 of that paper, where all 111 program stars (and the Sun) are listed together with their determined mean S-index values and the derived cycle periods whenever available.

The main lesson learned from the Mount Wilson HK chromospheric activity monitoring project (Baliunas et al. 1995) was that activity cycles are not an exclusive solar behavior but are relatively common among cool stars. Quite a few stars show activity cycles, and Baliunas et al. (1995) classify the quality of their derived cycle periods in categories of "excellent" (12 stars plus the Sun), "good" (eight stars), "fair" (15 stars) and "poor" (11 stars). For 51 stars S-index variability could be detected in the time series but without periodic behavior, and 14 stars were found to be "flat," i.e., showing no significant variability. Baliunas et al. (1995) conjecture that these latter stars may be Maunder minimum stars and represent the bottom cool star chromospheric activity.

Unfortunately, the term "Maunder minimum star" has not really been defined, and authors use this term differently. The Sun exhibited a low activity phase between $\approx 1645 - \approx 1715$, first pointed out by Spoerer (1890) and Maunder (1894). Although initially thought to be a likely observational artifact, the detailed investigation by Eddy (1976) provided convincing evidence for the reality of this extended solar minimum, which he dubbed "Maunder minimum." During this period the Sun showed few to no visible signs of activity, and any active regions were essentially absent. Moreover, coronal and chromospheric heating should have been minimal and therefore should not be related to a global dynamo, but rather indicate local processes. Later, Usoskin et al. (2007) studied sunspot numbers reconstructed from $^{14}$C data and identified a total of 27 solar "grand minima" episodes in their sunspot number reconstructions over the last 11300 years and conclude that the Sun spends about 17% of the time in such a "Maunder minimum" state. Thus, the Sun appears to shift from its state of "normal" cyclic activity, witnessed for the past 400 years, to an extended minimum state in a more or less regular fashion. Thus, such Maunder minimum eras of the Sun are apparently not at all that rare.

The only other star that has been claimed to have undergone a change from cyclic behavior to a Maunder minimum state appears to be HD166620 (Baum et al. 2022), and we discuss this specific star in Sect. 6.1. In the absence of such explicit changes, stars at the bottom of the observed activity range are proposed as Maunder minimum candidate stars, and we follow suite using this nomenclature. The number of such Maunder minimum candidate stars is under debate. While according to (Baliunas et al. 1995) the number of such objects in the Ca II HK Mount Wilson survey is quite substantial, for example, Järvinen & Strassmeier (2025) find extremely small fractions of such Maunder minimum candidate stars (two out of 13326) in their survey based on Ca II IRT data.

These early studies, however, gave rise to a number of new questions: first, cycles were shown not to be the only form in which cool stars express their magnetic activity in the time domain. Only a fraction of stars show truly solar-type cycles, according to the data presented by Baliunas et al. (1995). It remained open whether stars may exhibit other behavioral forms of magnetic activity, whether Baliunas et al. (1995) did miss any cycles, or whether the the anticipation of solar-type 11-year periods introduced a bias on cyclic stars with much shorter or longer periods. In this context we note the 121 day cycle of $\tau$ Boo (= HD120136), which was reported by Mittag et al. (2017) on the basis of TIGRE data, and a little later shown to be also present in the "historical" Mount Wilson data (Schmitt & Mittag 2017). Finally, one wonders how the strength and time domain behavior of magnetic activity coevolve with stellar evolution, and how does this coevolution depend, for example, on stellar mass, which produces differently deep convective envelopes along the main sequence. From the Sun it is well known that on a timescale of about nine cycles, weaker and stronger cycles occur, and a weak phase may result in a several decades long apparent absence of sunspot activity as was the Maunder minimum. So, one asks how stable or unstable are other cyclic stars, how the remnant activity of those stars in an apparent Maunder minimum state can be characterized, and which conclusions on the long-term prospects for solar activity can be drawn.

## 3. Observations and data analysis

In this paper we reconsider the "historical" Mount Wilson data (taken between 1966 and 1997) as well as new TIGRE data taken between 2013 and 2024. For our study we specifically focused on those 111 stars including the Sun presented by Baliunas et al. (1995), since for the large majority of this stellar sample HK observations going back to 1966 are available (see Fig. 1 in Baliunas et al. 1995). By combining the Mount Wilson and TIGRE data, one then obtains a sample covering a period of almost 60 years in time.

### 3.1. Mount-Wilson data

The H and K photometer, with which the bulk part of the Mount Wilson CaII H and K data were collected, is a four-channel photometer, consisting of a grating spectrometer, the output of which is sampled by a chopper (a rotating disk) and a photomultiplier; a detailed description of this instrument is presented by Vaughan et al. (1978). The system measures the counts in two bands with a width of 1.09 Å , centered on the CaII H and K lines as well as the counts in two continuum bands on either side of the H and K lines, i.e., the R band between 3991.067 Å - 4011.067 Å and the V band between 3891.067 Å - 3911.067 Å. Denoting the background corrected counts in these four bands by $N_V$, $N_K$, $N_H$, and $N_R$, the basic measurement is the so-called S-index defined as

$$S = \alpha \frac{N_H + N_K}{N_R + N_V}, \tag{1}$$

[1] The data can be downloaded from the site `https://dataverse.harvard.edu/dataverse/mwo_hk_project`.

[2] `https://nso.edu/data/historical-data/mount-wilson-observatory-hk-project/`

[3] `https://dataverse.harvard.edu/dataset.xhtml?persistentId=doi:10.7910/DVN/H85QPF`

where $\alpha$ is a normalization factor derived every night from observations of standard stars.

The only data available from the Mount Wilson HK Project are time series of the so determined S-indices. Usually (but not always), three observations of a given target were carried out consecutively. In this study we used the downloaded dataset, computed daily averages, and used the data scatter as a measure of uncertainty. We also note that a small fraction of the downloaded data must clearly be "wrong" for reasons that cannot be reconstructed in any individual case, perhaps because the wrong star was observed or that the spectral range of the HK photometer was incorrectly set. It is clear that any such "data editing" is subjective, but nothing is gained if obviously wrong data are used for further analysis.

### 3.2. TIGRE data

The TIGRE facility (TIGRE stands for "Telescopio Internacional de Guanajuato Robótico Espectroscópico") is a robotic spectroscopy telescope located in Guanajuato in central Mexico at the La Luz Observatory of the University of Guanajuato. Its 1.2 m telescope is fiber-coupled to an échelle spectrograph with a spectral resolving power exceeding 20 000 over the spectral range between 3800 Å and 8800 Å. In contrast to most other spectroscopic facilities, TIGRE operates fully robotically without any human intervention; a detailed description of the TIGRE hardware and performance is given by Schmitt et al. (2014) and an overview and summary of the first years of robotic spectroscopy with TIGRE is provided by González-Pérez et al. (2022). TIGRE's wide spectral coverage includes in particular the region of the Ca II H and K lines; hence, observations analogous to the Mount Wilson HK project can be conveniently carried out in addition to other spectroscopic monitoring projects.

Monitoring the Ca II emission of cool stars has been one of the prime objectives of TIGRE from the very beginning. A TIGRE S-index is calculated from the blaze-normalized and merged spectra. For simplicity a rectangular 1.0 Å band pass was used to represent the flux in the Ca II H and K line centers instead of the "classical" triangular band pass with a FWHM of 1.09 Å , which resulted from the rotating disk in the HK photometer. To compare these modern spectroscopy-based S-index measurements with the "historical" S-index measurements obtained using the HK photometer, any S-index measured with instrumentation other than the original HK photometer must first be transformed to the Mount Wilson scale. This is achieved by measuring the original set of calibration stars, which are known to be (more or less) constant, to enable a meaningful direct comparison. Without such a calibration, a comparison between the measurements taken by different instruments is at least questionable, and therefore great care was applied to derive this transformation. Mittag et al. (2016) derived the expression

$$S_{MWO} = (0.0360 \pm 0.0029) + (20.02 \pm 0.42)\, S_{TIGRE}, \qquad (2)$$

to convert our $S_{TIGRE}$ measurements to the scale of the "historical" Mount Wilson measurements. We note in passing that this calibration is consistent with an earlier calibration presented by Mittag et al. (2015). In this fashion the modern spectroscopic S-index measurements can be compared to the traditional HK photometer S-index measurements obtained by O. Wilson and his collaborators a few decades ago and thus calibrated and hence meaningful long-term time series can be constructed. We finally note that the TIGRE spectra cover the whole spectral range and, in particular, the region of H$\alpha$ line and the Ca II IRT; however, in this study we used only the information contained in the Ca II H and K lines.

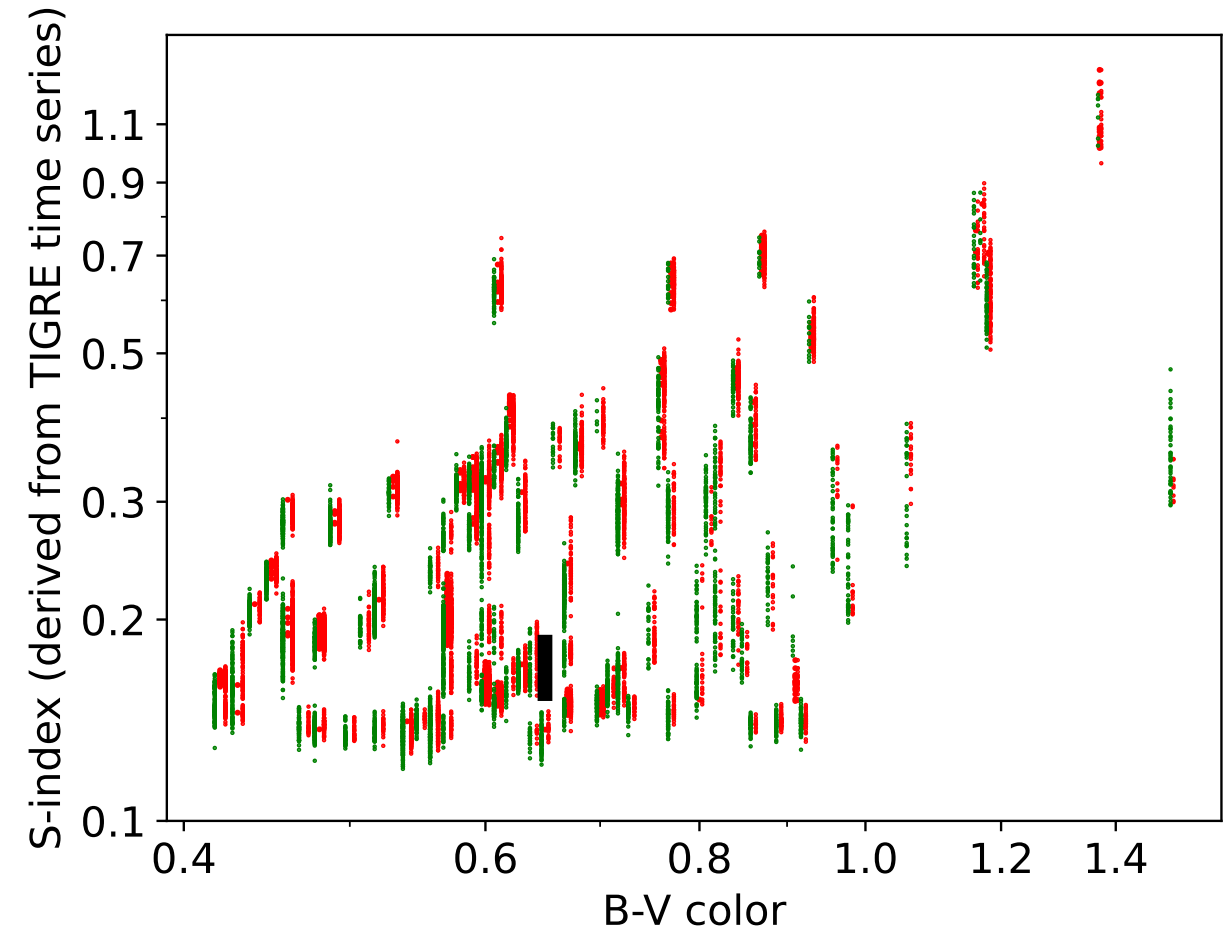

Fig. 1: S-index distributions of the stellar sample observed and derived by TIGRE. For better visibility the colored data points of a given star are slightly separated into the B-V color. The black bar indicates solar values (see text for details).

### 3.3. Stellar sample and S-index distribution

To provide an overview over the S-index values encountered in the Mount Wilson stellar sample, we compiled all of the measured TIGRE S-indices for our sample stars and plot the resulting data in Fig. 1 as a function of B-V color; in Fig. 1 we distinguish between data points with signal-to-noise ratios (S/N) > 150 (red points) and 150 > S/N > 100 (green points). Figure 1 shows that the bulk of our sample stars is in the color range 0.4 < B-V < 1.0 and that most of the sample S-indices are below a value of 0.4. Thus, the sample is somewhat concentrated toward lower activity levels. We also note that our solar data are not listed in Fig. 1; however, its observed range is indicated by the solid black line in Fig. 1. Figure 1 already shows that there are some stars with S-indices consistently lower than the solar values, and we return to this issue in Sect. 7.1.

### 3.4. TIGRE time line

With TIGRE we observed (almost) all of the stars in the Baliunas et al. (1995) sample in the period between 2013 to 2024. The number of observations taken for each star varies significantly. To provide an overview of the TIGRE data available for the Mount Wilson sample, we list in Table C.1 the annual number of spectra taken every year between 2013 to 2024 as well as the total number of spectra taken. While for one star (i.e., HD 120136 = $\tau$ Boo) more than a thousand spectra have been taken, others were not observed at all (these stars are discussed in Appendix A). Nonetheless, a few dozen spectra are typically available for each star.

### 3.5. Calibration accuracy

As to the calibration accuracy of the TIGRE S-index values, we refer to the extensive discussion by Mittag et al. (2016), who estimate a scatter between the calculated and the original reference values of $S_{MWO}$ of only 2.3% for TIGRE S-values (i.e., not transformed to the Mount Wilson scale) $S_{TIGRE} < 0.15$, but a scatter three times larger for the values above $S_{TIGRE} \approx 0.15$. To assess the validity of our calibration procedures in detail, we undertook a variety of further tests. First, we examined the accuracy of our cross calibration of the Mount Wilson and TIGRE S-indices by computing the mean yearly S-index of our sample stars as observed in the HK program and our TIGRE program. While naturally individual stars may increase or decrease their activity as measured by the S-index, one expects that in a sufficiently large stellar sample these effects ought to cancel out, and the sample mean should be close to constant. This is indeed the case. In Fig. 2 we plot the derived sample means (i.e., the mean of the stellar means obtained every year) as a function of time. Except for the first year (i.e., 1966), the sample means cluster around the value 0.26 between 1967 and 1995. In the remaining years, far fewer (35) stars were observed, which are a bit more active on average. To exclude the effects of different sample compositions, we further compared the mean S-index for the fixed stellar samples presented in Appendix D.

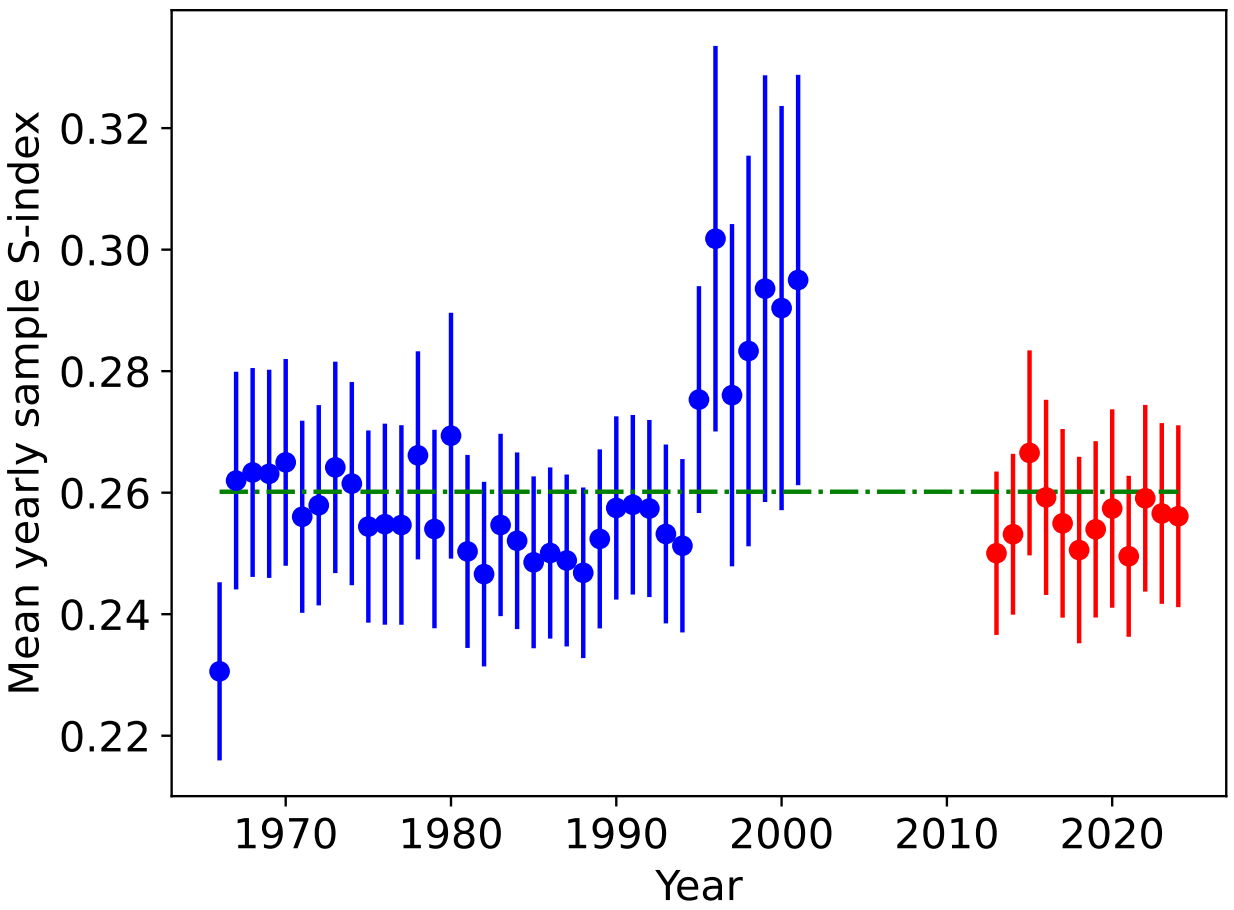

Fig. 2: Mean yearly S-index for the Mount Wilson stellar sample vs. time. The blue symbols indicate Mount Wilson data; and the red symbols indicate the S-indices derived from TIGRE observations.

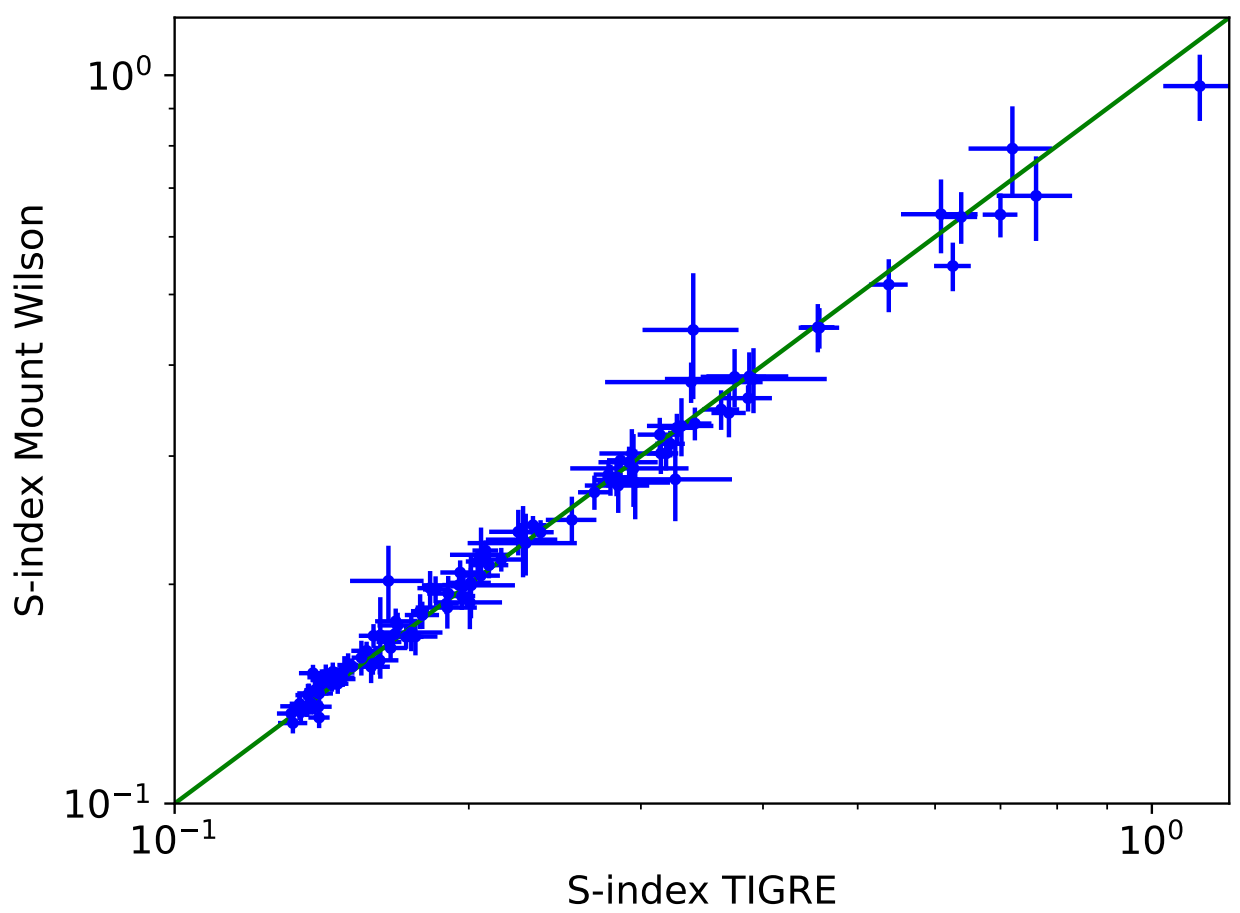

Fig. 3: Mean S-index for the TIGRE stellar sample vs. Mount Wilson mean values. The green line indicates identity (see text for details).

Next, we ask how the mean S-indices derived from the "historical" Mount Wilson observations compare to the mean S-indices derived from our "new" TIGRE observations on a star-by-star basis. This comparison is shown in Fig. 3, where we show for all common sample stars the mean S-indices derived from the Mount Wilson observations versus those derived from TIGRE. Figure 3 demonstrates that – with some exceptions – the respective S-indices are within the scatter of the respective datasets. This finding already has a scientific significance that we address later in Sect. 7; it appears that none of our sample stars has undergone a dramatic change in its mean activity. From these two tests, therefore, we draw three conclusions. First, our cross calibration is correct; second, no systematic errors have been introduced; and third, the "historical" Mount Wilson data can be meaningfully compared to our "new" TIGRE data.

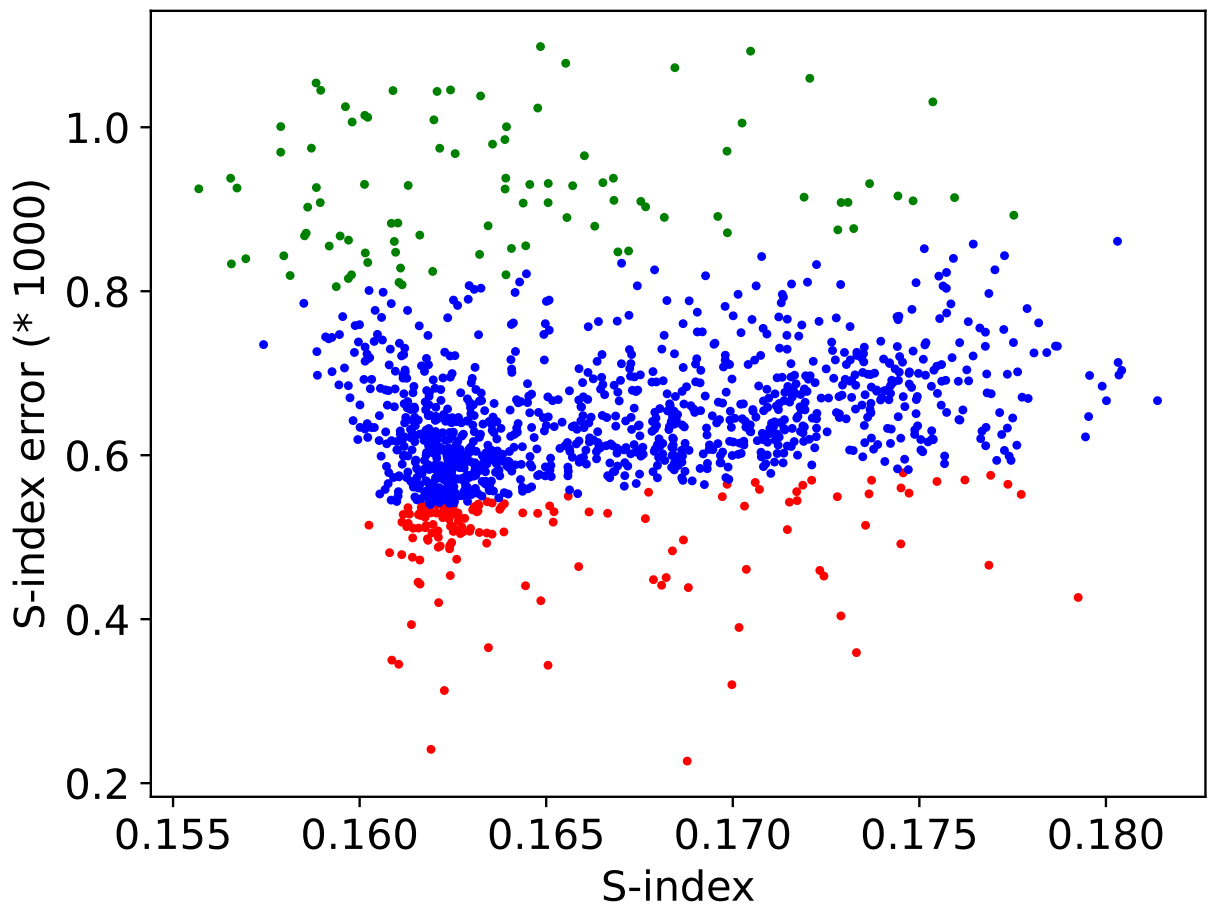

Fig. 4: S-index errors vs. the derived solar S-indices for all TIGRE data; red data points refer to data with a S/N > 300, blue data points to data with 200 < S/N < 300 and green data points to measurements with 150 < S/N < 200. Measurements with a formal error above 0.001 were not considered in the subsequent analysis ; see text for details.

## 4. The Sun as a star

The Mount Wilson stellar sample exclusively consists of cool solar-like stars. It is therefore imperative to check what the Sun looks like when observed in the same fashion as our whole stellar sample. Egeland et al. (2017) present a long-term time series of solar S-index values covering the period between 1966 - 2016 derived from various sources, i.e., from direct measurements with the Mount Wilson HK photometers, with which the Moon was observed, from the solar-stellar spectrograph (SSS), which observed the Sun with the same spectrograph as used for stars, and from direct solar measurements of the K index, transformed to the S-index scale. In this context it is useful to remember that only a relatively small number of data points presented by Egeland et al. (2017) have actually been obtained using the HK photometers and we refer to Egeland et al. (2017) for a detailed discussion of this specific time series. In our study we focus solely on Mt. Wilson and TIGRE data, as both datasets are well-calibrated by the same large set of calibration stars, while several other attempts to obtain S-values have only secondary calibrations, where offsets in the S-index scale cannot be ruled out. Finally, as shown by Vanden Broeck et al. (2024), there is a very good correlation between the (full-disk) solar S-index as measured by TIGRE and the fractional area of chromospherically bright regions on the Sun, a finding that illustrates the empirical importance of the S-index and its temporal variations, and naturally one expects the same phenomena to occur in solar-like stars.

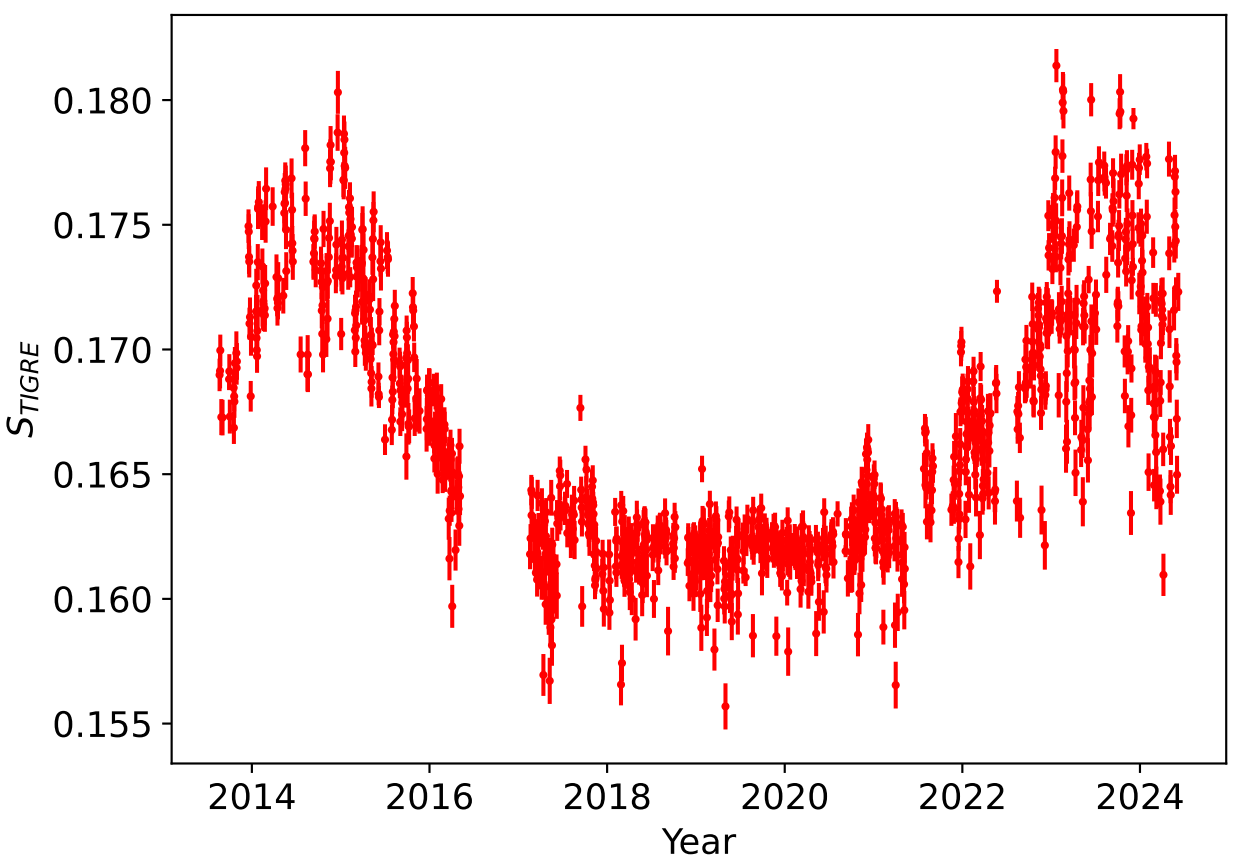

Fig. 5: Solar S-index vs. time measured by TIGRE from 2013 until 2024; see text for details.

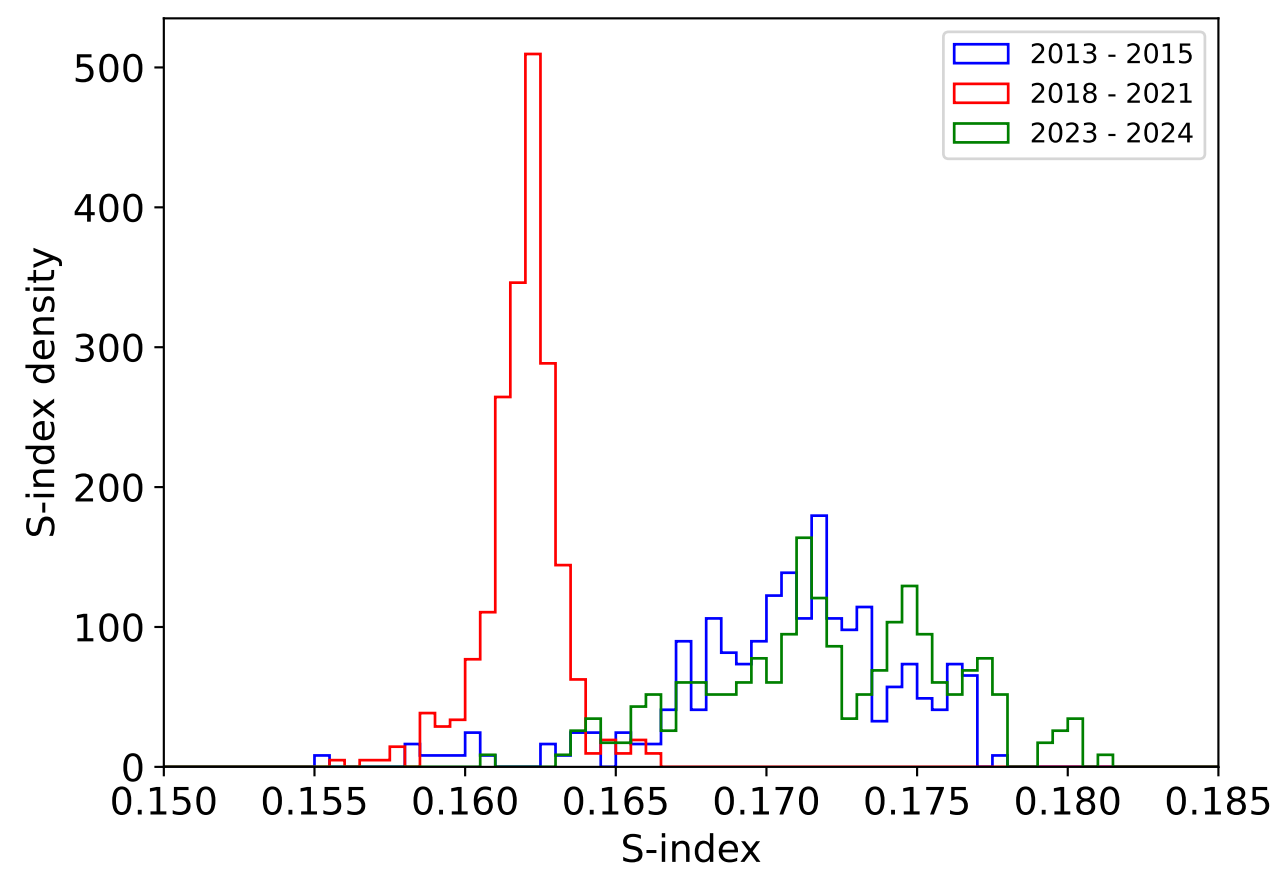

Fig. 6: Histogram representation of the solar S-index variability; red histogram refers to the solar minimum between 2015-2018, the blue and green histograms to the solar maximum between the years 2013-2015 and 2023 - 2024, respectively.

### 4.1. S-index and signal-to-noise

As emphasized by Egeland et al. (2017) and of course also reemphasized by us, to place any S-index measurement on the Mount Wilson scale, it is essential to use some calibration proxy; otherwise, such comparisons are not meaningful. For these reasons S-index determinations for the Sun were considered important also for TIGRE, and more than 5000 individual solar measurements were taken between 2013 - 2024. For these measurements we chose the solar light reflected off the Moon. Typically, four measurements were performed: the first was a test measurement to determine the required exposure time, followed by three science measurements. However, in some cases, more – or occasionally fewer – measurements are available. The first test measurement usually has a lower S/N than the subsequent science measurements. We averaged the lunar observations of a given night, but excluded all measurements from the averaging process that deviate by more than 0.01 from the mean. For each measurement we compute a formal signal-to-noise ratio of the S-index and plot in Fig. 4 the obtained TIGRE S-index measurements and their derived formal errors as a function of the signal-to-noise ratio. One recognizes from Fig. 4 that – not surprisingly – the error increases with decreasing S/N and that lower S/N measurements extend to lower S-indices. For our analysis we therefore chose to ignore all data points with a formal error above 0.001 and thus end up with 1287 valid measurements of the solar S-index between 2013 and 2024 taken with the same instrumentation that was used for the stellar S-index measurements. None of our TIGRE measurements drops below an S-index value of 0.155, which therefore appears to represent an empirical lower limit to the observed solar S-index variability at least in the time frame 2013 - 2024.

### 4.2. Solar S-index time series

The resulting solar S-index time series homogeneously derived from the TIGRE data is plotted in Fig. 5, which shows that the solar maxima near 2015 and 2023/2024 have been well covered, both exhibiting a higher degree of variability, but also the minimal Sun between the years 2018 and 2020 is well covered by our TIGRE time series. In Fig. 6 we show a histogram representation of our solar S-index time series. The S-index distributions for the solar maximum are very similar for the maxima in 2015 and 2024, while the S-index distribution for the solar minimum is quite distinct and much narrower; both Figs. 5 and 6 demonstrate that the solar maximum S-values encountered during the maxima in 2015 and 2023-2024 are quite similar, which is also reflected in the respective Sun spot data available, for example, through the Solar Influences Data analysis Center at the Royal Observatory of Belgium.

Our TIGRE solar S-index measurements cover the range between 0.155 to 0.18, which then is the solar S-index range in the period 2013 - 2024. The lower value lies at the bottom of the observed stellar range. Also, during the extended solar minimum in 2008, we measured similar values. However, according to Egeland et al. (2017), earlier solar maxima starting in the 1950s reached larger S-values. In Fig. 7 we compare our TIGRE solar S-index measurements with the solar S-index measurements presented by Egeland et al. (2017); we also show TIGRE data obtained from 2008 to 2009 (black data points) during the telescope's test phase in Hamburg. However, we do point out that spectra taken with reflected moonlight and reflected atmosphere were used and a different calibration was applied (for details see Schröder et al. 2012). The green data points in Fig. 7 refer to data taken with the SSS instrument, and the blue data points are measurements obtained with the HKP1/2 instruments. For a detailed discussion, we refer again to Egeland et al. (2017). Note that between the years 1979 and 1992 no direct S-index measurements are available, rather S-indices are inferred from solar K line measurements and therefore left out in Fig. 7. We also note in this context that Baliunas et al. (1995) report solar S-indices reaching up to 0.20, yet it appears that a revised calibration of the HK data as presented by Egeland et al. (2017) lowers these values. As far as the direct S-index measurements are concerned, there is good agreement between HKP and SSS measurements and between SSS and TIGRE measurements in the regions of overlap. We therefore conclude that there do not appear any serious cross-calibration issues.

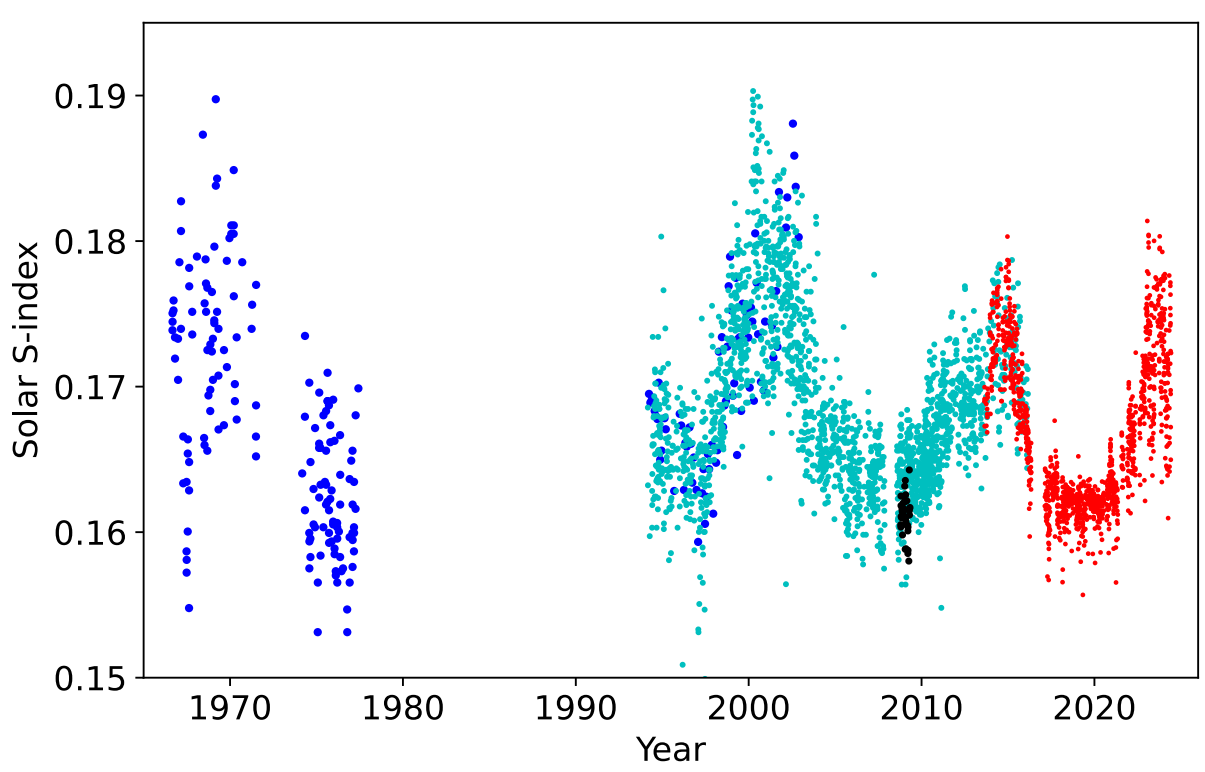

Fig. 7: Comparison of TIGRE derived S-index measurements presented by Egeland et al. (2017). Red data points are TIGRE data, black data points are not fully calibrated TIGRE data taken during an earlier test phase in Hamburg, and cyan data points are SSS measurements, blue data points are HKP1/2 measurements; see text for details.

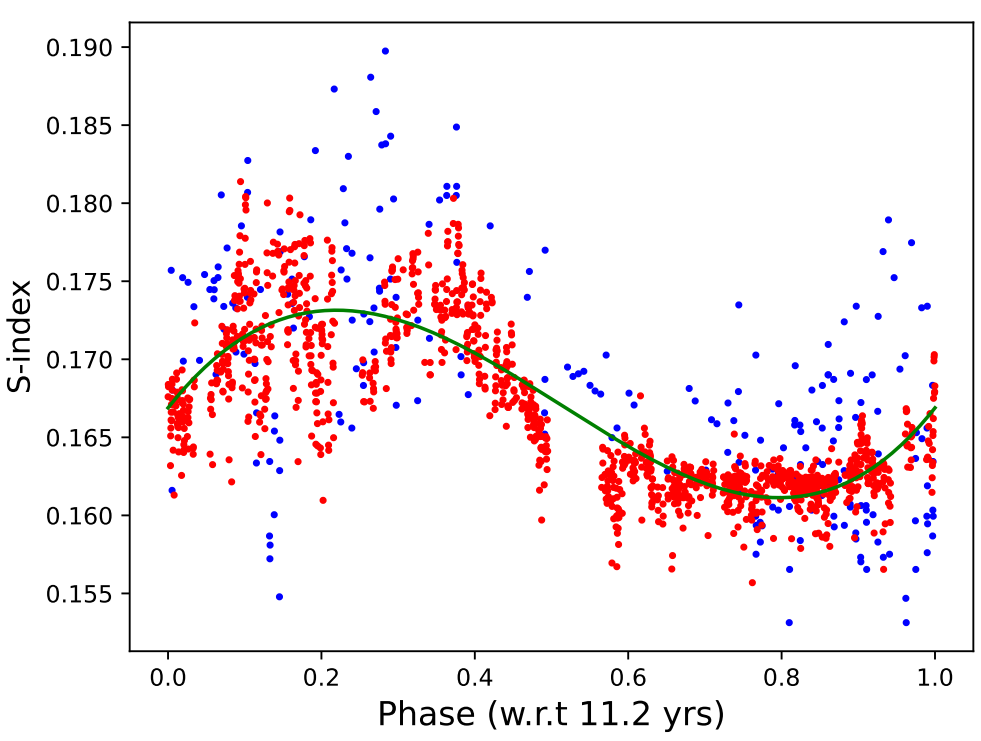

Fig. 8: S-index light curve of the Sun using only Mount Wilson HK and TIGRE data, folded with the lowest variance period of 11.2 years. The blue data points indicate HK data, and the red data points indicate TIGRE data. The green curve represents a polynomial fit to the phased curve as described in Sect. 5.1.1.

Finally, and in anticipation of our stellar S-index work and as a sanity check, we phase-folded the solar S-index data, using only the actual Mount Wilson HK data and our own TIGRE data. For any given trial period, we phase-folded and fit the resulting phased S-index curve with a polynomial as described in detail in Sect. 5.1.1. The S-index light curve folded with a period of 11.2 years shows the least amount of dispersion and is shown in Fig. 8. While obviously quite a bit of variability at a given cycle phase exists, our procedure correctly recovers the well known solar cycle period.

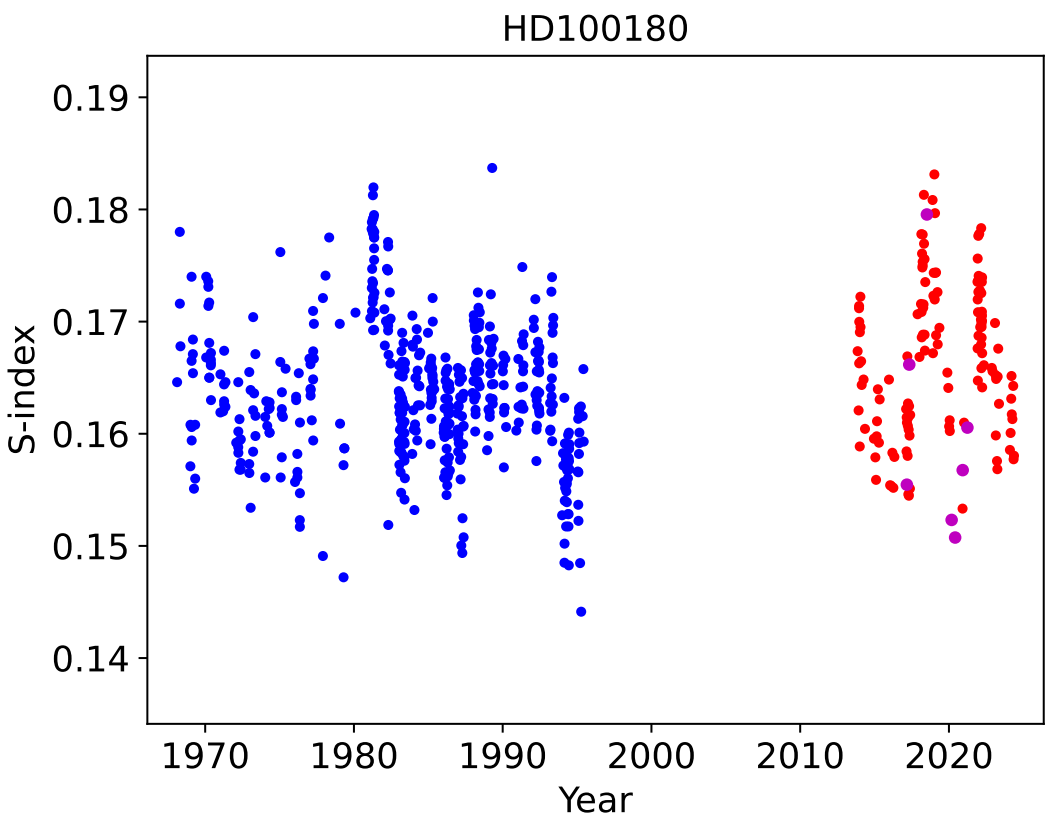

Fig. 9: Joint Mount Wilson (blue data points) and TIGRE S-index light curves (red data points with S/N > 100) for HD 100180. The magenta data indicate S/N in the 50 < S/N < 100 range (see text for details).

## 5. Construction of long-term stellar Ca II time series

For every star in the Mount Wilson sample, we assembled all TIGRE measurements taken from 2013, when TIGRE telescope online in Mexico, until May 2024. TIGRE data reduction was performed with a special pipeline; for a detailed description, see Hempelmann et al. (2016) and Mittag et al. (2016). A special feature of this pipeline is that it includes an automatic determination of S-indices, which were used in this paper together with the calibration described in Sect. 3.2. We normally used only data where the individual S-index measurements have a S/N of more than 100; however, in the light curves we also considered measurements with S-indices with a S/N of more than 50.

### *5.1. Joining Mount Wilson and TIGRE data*

#### 5.1.1. A case study of HD 100180

To demonstrate our analysis procedures, let us consider as an example the case of HD100180 (= 88 Leo), an F7V star with a "fair" cycle as determined by Baliunas et al. (1995) who derived a cycle length of 3.56 yrs. The S-index data used for the analysis are displayed in Fig. 9, the individual data analysis steps are illustrated in the four panels of Fig. 10. Note that our periodograms are always generalized Lomb-Scargle (GLS) periodograms, as described by Ferraz-Mello (1981) and Zechmeister & Kürster (2009); for an in-depth discussion of the Lomb-Scargle periodograms, we refer the reader to VanderPlas (2018).

In the particular case of HD100180, the TIGRE data capture the 3.56 yr period derived by Baliunas et al. (1995) exceedingly well and provide a very high power peak, while the Mount Wilson period is less clear (after all, the cycle was only classified as "fair" at the time). A weaker secondary peak appears near a period of ten years. This could either indicate a second, longer cycle, as suggested by Baliunas et al. (1995), who derived 12.9 yrs, or it could be related to aliasing issues. However, the period ratio is 2.7, rather than the expected 2 for aliased periods.

Given the interpretational difficulties encountered with a GLS periodogram with non-sinusoidal variations, we decided to base our final classification on the phased S-index light curves using a phase dispersion minimization (PDM) approach. While the original PDM approach by Stellingwerf (1978) uses some ar-bitrary number of phase bins that can be specified ahead of time, we described the resulting phased light curve with a fourth order polynomial $P_4(x)$ in the range $0 \leq x \leq 1$, such that $P_4(0) = P_4(1)$ and $P'_4(0) = P'_4(1)$ so that the resulting fitted phased light curve is continuous and differentiable. Clearly, such a polynomial $P_4$ has three free parameters, which must be fitted from the data. Obviously, the choice of a fourth order polynomial is arbitrary, but polynomials of higher orders often result in instabilities and several phased light curve maxima, which appear unphysical.

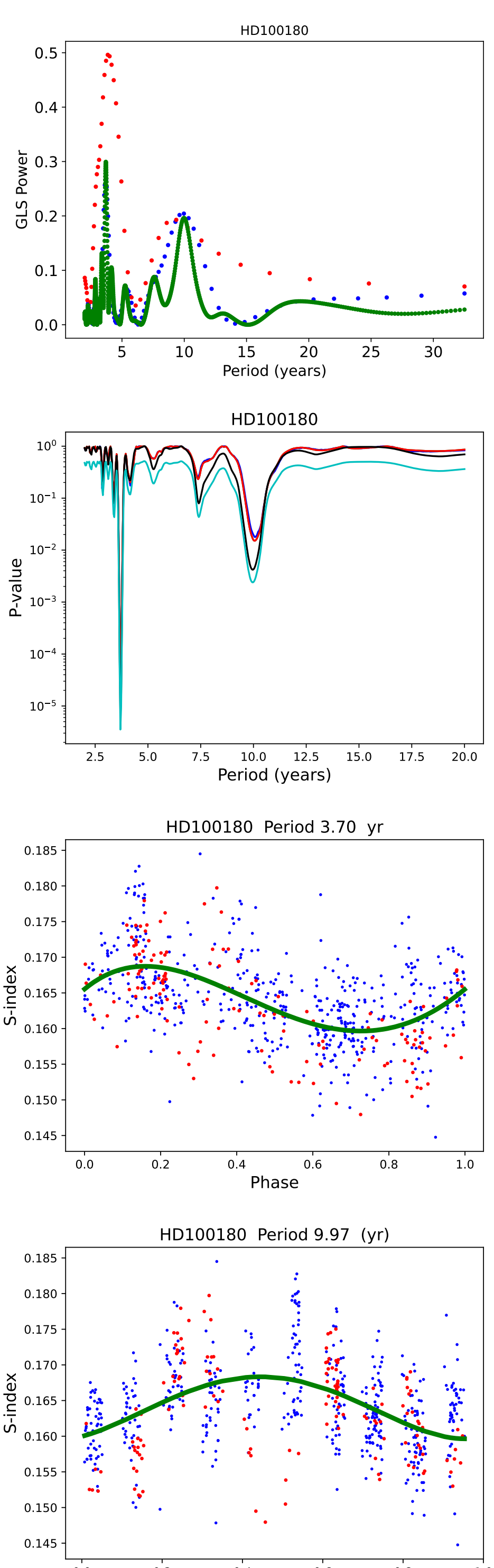

Fig. 10: Demonstration of analysis procedures. Top: GLS periodograms for the individual datasets (blue and red points) as well as joined datasets (green points) taken from Fig. 9. Second from top: Resulting p-values. Lower panels: Data shown folded with the most significant periods. The green curves represent polynomials constructed as described in Sect. 5.1.1 (see text for details).

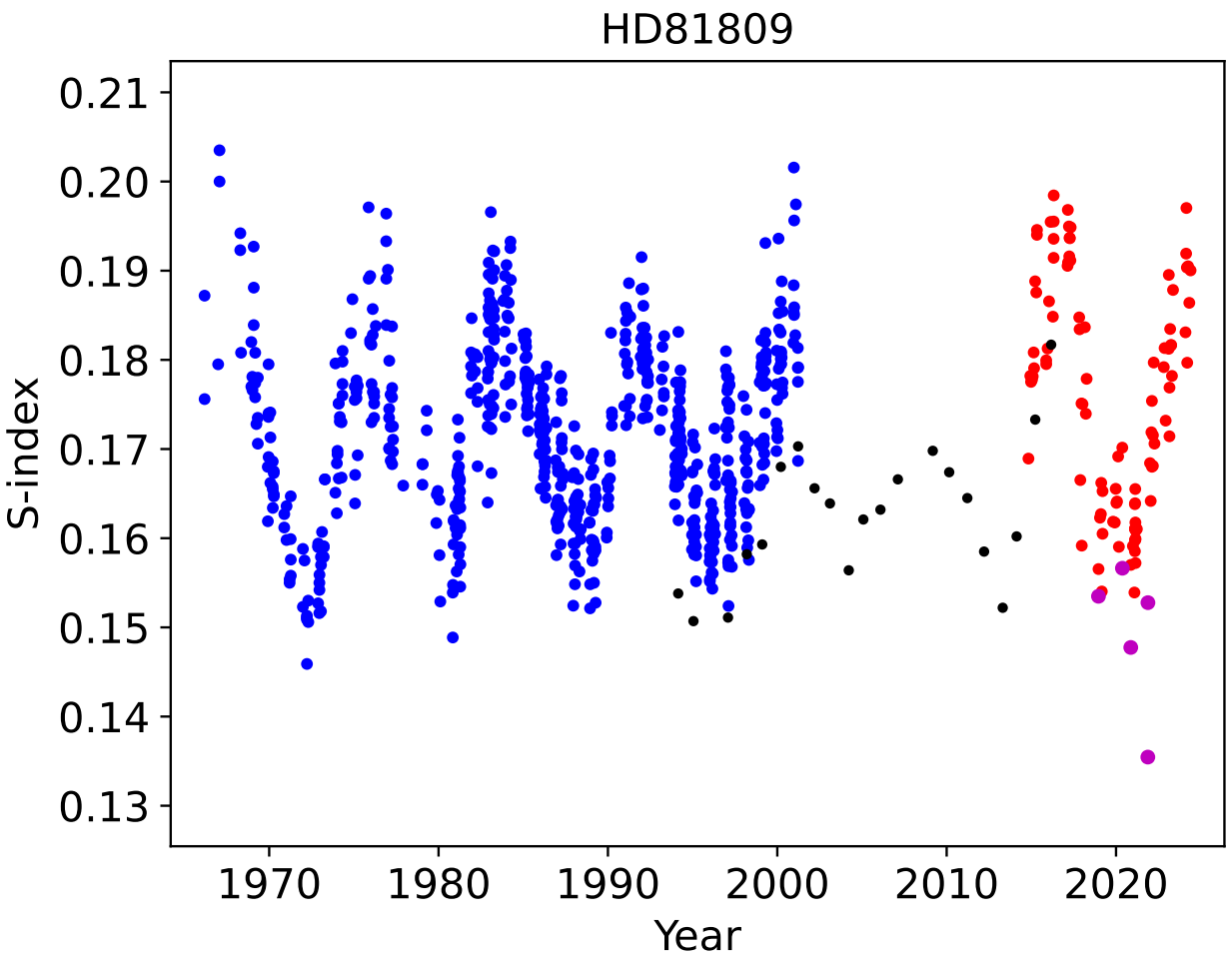

Fig. 11: S-index variability for HD 81809 between 1966 and 2024. The blue data points refer to Mount Wilson observations, the black data points represent seasonal medians obtained with SSS, and the red and magenta data points refer to TIGRE observations.

We then computed the dispersion of the phased data points with respect to the best-fit phased light curve and compare this value to the dispersion of the phased light curve with respect to a constant value. A variety of tests exists, which can assess whether introducing the polynomial $P_4$ provides a significant dispersion reduction; for a detailed description of these tests, we refer the reader to Conover et al. (1981). Here, we used the *python* implementations of the F-test, the Bartlett test, the Levene test, and the Fligner test, which all use somewhat different test statistics (see Conover et al. 1981). We find that, naturally, these tests provide somewhat different significance levels; however, our results are essentially independent of the specific test chosen, and we used the p-values from the highest significance test. The specific example of HD100180 is shown in the second panel from top in Fig. 10, where the significance of the variance reduction versus chosen folding period is shown for the various tests. One recognizes that all tests suggest a very significant periodicity at 3.7 yr and another – less significant – periodicity at 9.97 yr.

In the lowest two panels of Fig. 10, we then show the totality of S-index data points folded with the peak power period of 3.70 yr and of 9.97 yr with the individual data points color-coded as before. The green solid line in the panels is not a physical fit. Rather, it merely shows the best-fit polynomial $P_4$ constructed the way described above. Figure 10 (second panel from bottom) shows that the Mount Wilson and TIGRE data points are clustered quite uniformly around the solid line suggesting that the 3.7 yr cycle of HD100180 is stable over (at least) 60 years. Similar considerations apply to the data when folded with the second period of 9.97 y, except that at the expected cycle maximum the TIGRE data is low, while the rise to and the decay from maximum is well captured. In the same fashion, as demonstrated in Fig. 10 for the case of HD100108, all other sample stars were analyzed.

#### 5.1.2. A case study of HD 81809

Figure 9 shows a gap in the S-index data between the end of the Mount Wilson observations and the start of the TIGRE observations. As previously mentioned, Ca II H and K monitoring data are also available from the SSS at the Lowell Observatory (Hall & Lockwood 1995), which fully or partially cover the time between 1994 and 2016. Radick et al. (2018) published time series for 72 stars in the form of seasonal medians for these objects, 40 of which overlap with our stellar sample. In Table C.1 those of our sample stars with SSS data available are marked by a ”Y.” The SSS coverage for the different sample stars significantly varies (cf. Table 3 in Radick et al. 2018). One of the best cases is the star HD 81809, whose cycle was already discovered by Wilson (1978) and whose X-ray emission also shows a very clear cyclic behavior (Favata et al. 2008; Robrade et al. 2012) with the same period as observed in Ca II H and K. In Fig. 11 we show the resulting S-index curve for HD 81809, covering the full time span between 1966 and 2024. The cyclic variability is very obvious, eight maxima have been covered so far and the cycle appears to be very stable during this time. Also, the common data between TIGRE and SSS shows that the calibration between the two instruments appears to be consistent. Note that the SSS data (i.e., the black data points) are seasonal medians and thus can not be directly compared to the HK and TIGRE data. Yet, it appears that the two maxima covered by SSS are lower than the other six observed maxima. Whether or not this indicates longer term cycles, we can only speculate.

We must recall in this context that the case of HD 81809 with respect to SSS data coverage is exceptional. For most other stars with available SSS data, the coverage is much sparser, and for the majority of our sample stars no SSS data are available at all. Thus, to provide a homogeneous analysis for all stars, we neglected any SSS data in our variability analysis and merely show the SSS data whenever available for demonstration purposes.

### *5.2. Ca II time series over half a century*

In the following sections, we provide a detailed comparison between the TIGRE results and the prior Mount Wilson monitoring. We follow the classification given by Baliunas et al. (1995) and separately discuss the stars with a cycle period classification ”excellent” , ”good,” ”fair,” and ”poor.” We also consider the ”flat” stars without any detectable variability as well as the variable stars without any discernible period. However, in contrast to Baliunas et al. (1995) we use a quantitative classification criterion as described in Sect. 5.1.1: we consider all stars with a cycle p-value smaller than $10^{-5}$ as ”excellent,” those with p-values between $10^{-5}$ and $10^{-4}$ as ”good,” those with p-values between $10^{-4}$ and $10^{-3}$ as ”fair,” and those with p-values between $10^{-3}$ and $10^{-2}$ as ”poor”; stars with lower p-values are considered noncyclic. We discuss the various groups of stars in Sect. 6 and, for better readability of our paper, we provide the relevant tables and figures in the appendices or only online.

## 6. How stable are stellar magnetic cycles?

### *6.1. Stars with ”excellent” cycles*

According to our classification scheme outlined in Sect. 5.2, we find 32 stars (including the Sun) with ”excellent” cycles. In Table E.1 we provide a list of these stars together with our modeling results. In Fig. F.1 we provide the phased S-index light curves of exemplary three ”excellent stars,” which were not classified as "excellent" by Baliunas et al. (1995) [4]; their derived periods are listed in Table E.1 and additionally on top each figure. Each figure contains the Mount Wilson data (blue data points), the TIGRE data (red data points), and a mean phase curve (solid green line) derived as described in Sect. 5.1.1. In these cases where SSS data are available, they are also shown as black data points, but these data were not used in our period derivations to ensure a maximally homogeneous data analysis. We further note that in most cases the cyclic variability is readily visible ”by eye.” The phase coverage of the TIGRE observations (which cover only ten years) is pretty good for most stars. The only exception is HD 166620 whose suggested cycle period is 16.72 yr. The p-values of the derived periods are – naturally – very high, the ”old” and ”new” periods are very similar and the mean cycle range (i.e., the difference between cycle maximum and minimum divided by the mean) is between 15% - 30% in all cases.

In Table E.1 we also list (in the column denoted by ”MW”) the results derived by Baliunas et al. (1995) from the Mount Wilson data alone as well as the results derived by Oláh et al. (2016) for the 29 stars analyzed by these authors (column O2016). An inspection of Table E.1 reveals that all stars classified by Baliunas et al. (1995) as ”excellent” are also considered ”excellent” in our classification scheme. Moreover, an inspection of the relevant S-index light curves plotted in Fig. F.1 shows that for all ”excellent” stars (according to Baliunas et al. 1995) the S-index variability seen in the TIGRE data is consistent with the cyclic variability derived about 40 years earlier. To put it differently, the cyclic variability pattern observed in these stars appears to be stable over a time period of 60 years, and none has entered an actual Maunder minimum episode.

The only marginal case in this respect appears to be the star HD166620 with a period of 16.3 years, which, however, is longer than the time span covered by the TIGRE observations. While this period is very obvious in the Mount Wilson data, TIGRE unfortunately missed the expected cycle maximum, yet the TIGRE data is still consistent with the Mount Wilson observed cyclic variability. In this context we note that HD166620 has been proposed by Baum et al. (2022) as a Maunder Minimum candidate star that has changed from a cyclic to a constant behavior; we just remark that our TIGRE data are not consistent with such an interpretation.

Before entering a detailed comparison, we present a comparison of a subset of twenty nine stars analyzed by (Oláh et al. 2016) with a time-frequency analysis. In this set (Oláh et al. 2016) find thirteen stars with a dominant cycle period, which agrees very well with our periods derived in a different fashion. For seventeen stars in their sample, (Oláh et al. 2016) find ”complex” cycles with several periods or a complex phase behavior. For thirteen out of these seventeen stars, our analysis finds significant cycle periods, and in many (but not all) cases our periods agree reasonably well with those presented by (Oláh et al. 2016).

Comparing our ”excellent” stars as listed in Table E.1, we note that only 13 received the same classification by Baliunas et al. (1995), yet seven had been classified by Baliunas et al. (1995) only as ”good,” six as ”fair,” one as ”poor” and five had no detected cyclic variability. We now examine those stars in detail in the following and address the ”only good” stars first, i.e., the stars HD 78366, HD 114710, HD 115404,

[4] Phased S-index light curves for all stars listed in Table E.1 with detected cycles are available online.

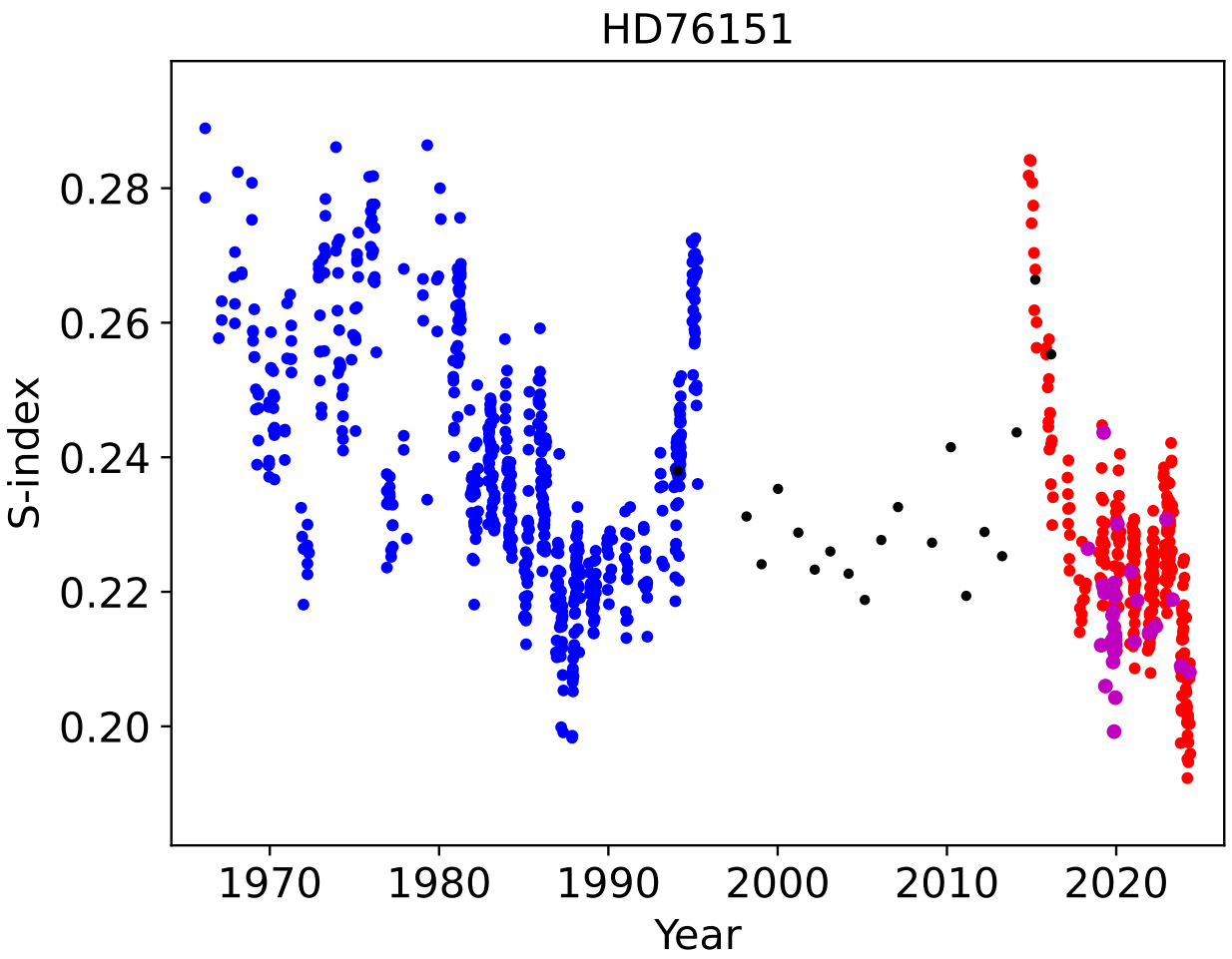

Fig. 12: S-index light curve for HD76151 between 1966 and 2024. The blue data points indicate Mount Wilson data, and the black data points indicate SSS seasonal medians. The red and magenta points indicate TIGRE data (see text for details).

HD 149661, HD 156026, HD 201092, and HD 219834A. In the case of HD 78366, we notice significant seasonal variability, and the period of 12.2 yr implies that the TIGRE data do not cover the full cycle. Yet, the cyclic character of the light curve is very obvious from the Mount Wilson data, and the TIGRE data are consistent, whereas the SSS data are low but provide only extremely limited phase coverage. Similar remarks apply to HD 114710. Again with a period of 17.58 yr, the TIGRE data do not cover a full cycle missing out a third of the phase interval; nevertheless, the expected cycle maximum is very well covered. For HD 115404 we find a period of 10.9 y consistent the findings of Oláh et al. (2016), while the TIGRE data alone are inconclusive. For HD 149661 we find a clear period of 11.76 y with some multi-period behavior, our secondary period is very close to the period reported by Oláh et al. (2016). For the star HD 156026, "our" period of 14.83 yr fits both the Mount Wilson and TIGRE data very well, and the same applies to HD 201092 and to HD 219834A.

Next, we turn to the "only fair" stars according to Baliunas et al. (1995), i.e., the stars HD 76151, HD 100180, HD 154417, HD 161239, HD165341A, and HD 190406. HD 100180 has been discussed already. For HD 76151 the Mount Wilson and TIGRE data agree well with a common long period; however, the S-index light curve of HD 76151 looks a little odd and shorter periods might be present. In Fig. 12 we show its S-index variation between 1966 and 2025, which in particular show the end of the Mount Wilson data (blue data points), the SSS data (season medians, black data points) and the TIGRE observations (red data points). Inspection of Fig. 12 shows that during the first two TIGRE observing seasons (i.e., 2014/2015 and 2015/2016) the S-index was constantly decreasing from around 0.285 at the end of 2014 to around 0.232 in the second quarter of 2016. This decrease is definitely not an artifact, it is also discernible in the SSS data. Also, the SSS data point of the 2013–2014 season is only moderately enhanced, the S-index of the prior season is low. This unusual high flux episode of HD 76151 appears to have lasted about two to three years. In the case of HD 154417 the Mount Wilson data show a clear periodic signal and the TIGRE data are consistent, the same applies to HD 161239, HD165341A, and HD 190406. Finally, in the case of HD 76572, Mount Wilson and TIGRE data agree well with a long period of 16.4 y.

Finally, we consider the five stars without any cycle determination on the basis of the Mount Wilson data. For HD 10780 substantial seasonal variability exists; nevertheless, TIGRE samples the 9.69 yr cycle exceedingly well. Similar remarks apply to HD 22049 with a cycle period of 10.77 years, which again appears in both the Mount Wilson and TIGRE time series. Further details on HD 22049 are given in Sect. 6.6.2 and by Fuhrmeister et al. (2023). HD 72905 with a cycle period of 11.9 yr is again sampled quite well, and a periodic variability is present. The case of HD 101501 is more complicated. Inspection of the p-values and the GLS periodogram shows that a periodicity of 4.5 yrs may also be present in the data, which one also recognizes in the phase-folded light curves. Thus, we conclude that the true period(s) of HD 101501 are uncertain at the moment. In the case of HD 185144 the period of 6.28 yr is very significantly present in both the Mount Wilson and TIGRE data.

### 6.2. Stars with "good" cycles

In Table E.1 we also list three stars, i.e., HD 3651, HD 129333, and HD 187691, which we classify as having "good" cycles, the three exemplary joint Mount Wilson-TIGRE phased light curves are displayed in Fig. F.2; only for HD 187691 exists some SSS data and inspection of the light curves shows that in the case of HD 187691 the period is driven by the Mount Wilson data, while the TIGRE data prefer half this period and the SSS data are rather flat.

### 6.3. Stars with "fair" cycles

Next, we turn to the eight stars with a "fair" cycle classification, which are also listed in Table E.1, i.e., HD 6920, HD 26913, HD 26923, HD 88737, HD 115043, HD 155885, HD 176051, and HD 224930. Three exemplary phased light curves are shown in Fig. F.3. Only one star (HD 26913) also received a "fair" cycle classification by Baliunas et al. (1995); two were classified as "poor" (HD 155885 and HD 224930); and the rest had either questionable or no periods at all. The phased light curves displayed in Fig. F.3 look quite convincing. In the case of HD 115043, the short cycle of 4.27 yr has been sampled many times; however, other periods may also be present.

### 6.4. Stars with "poor" cycles

Finally, we turn to fifteen stars the cycles of which we classify as "poor" as listed in Table E.1. Three exemplary light curves are again shown in Fig. F.4. Four of those stars (HD 18256, HD 20630, HD 157856, and HD 190007) were assigned "fair" cycles by Baliunas et al. (1995), two of them (HD 82443 and HD 188512) were assigned "poor" cycles by Baliunas et al. (1995), while the rest had no cycles associated with them. The exemplary phase curves shown in Fig. F.4 look reasonable, a comparison with the periods derived by Baliunas et al. (1995) shows agreement except for HD 82443, which exhibits a rather unusual S-index light curve.

### 6.5. Stars with multiple period cycles

In a few cases, the resulting S-index light curves are consistent with more than one period. One of these cases is the star HD 100180 already discussed in detail in Sect. 5.1.1. Both the GLS power and p-value versus period curves (cf. Fig. 9) show two clear peaks, one main peak at a period of 3.7 yr and a secondary peak with lower significance at a period of 9.97 yr. In Fig. 11 we show the Mount Wilson and TIGRE data folded with this secondary period of 9.97 yr, which also results in a reasonably looking phased S-index light curve. While in the case of HD 100180 the best-fit periods have different classifications, there are seven cases listed in Table E.1, where two or sometimes even more periods have the same level of significance, i.e., the cases of HD 22049, HD 78366, HD 82443, HD 152391, HD 154417, HD 156026, and HD 185144. Three cases are also contained in the sample by Oláh et al. (2016) all in their "C" category. We examined the p-value versus period curves for all these cases, a complicated task, particularly for HD 82443, whose p-value versus period curve contains six minima. Overall, we can state that the secondary or tertiary periods listed in Table E.1 do not differ that much, but it is difficult to arrive at firm conclusions with the available data. Clearly these stars are prime candidates for differential rotation but this would normally require a higher cadence and longer monitoring.

### 6.6. Remarks on individual stars

We conclude this section by briefly listing well-known stars, which have already been studied in larger detail by other authors and are therefore not discussed in detail in this context.

#### 6.6.1. $\tau$ Boo (= HD 120136)

The star $\tau$ Boo is the object with the largest number of individual TIGRE observations (see Table C.1). These observations led to the discovery of a rather short activity cycle with a period of about 122 d Mittag et al. (2017); a similar period was also reported by Mengel et al. (2016). A subsequent reanalysis of the Mount Wilson data presented by Schmitt & Mittag (2017) showed that this cycle was actually already present in these data but went unrecognized at the time.

#### 6.6.2. $\epsilon$ Eri

Fuhrmeister et al. (2023) used TIGRE data in a multiwavelength analysis of the nearby young star $\epsilon$ Eri (HD 22049). In their study, they analyzed both chromospheric and X-ray monitoring data on this star and find correlated cyclic behavior in Ca II H and K and X-rays on a timescale of 3 yr, as well as evidence for longer chromospheric cycles at 11 yr and 34 yr. In our analysis the 3 yr period is clearly evident in the TIGRE data but not in the Mount Wilson data. However, joining TIGRE and Mount Wilson observations yields a very strong peak at 10.8 yr and makes $\epsilon$ Eri an "excellent" cycle star.

## 7. Activity of noncyclic stars and cyclic stars in comparison

### 7.1. S-index distributions and Maunder minimum candidate stars

About half our sample stars show no detectable cyclic variability, which raises the question whether these stars are also different in other respects. Therefore, we first consider activity as measured through the S-index for the noncyclic stars, the stars with excellent cycles and the rest, i.e., the stars with less significant cycles, which we lump into one group. In Fig. 13 we show a histogram of the observed mean S-index distributions for these three groups. As is obvious from Fig. 13, the activity distributions of the cyclic stars do not differ a lot, yet it is clear that the vast majority of the noncyclic stars have very low activity, notwithstanding some noncyclic (red) stars with larger activity. What is especially noteworthy about Fig. 13 is that a group of stars exist that have mean S-indices below solar values and no detectable cyclic variations.

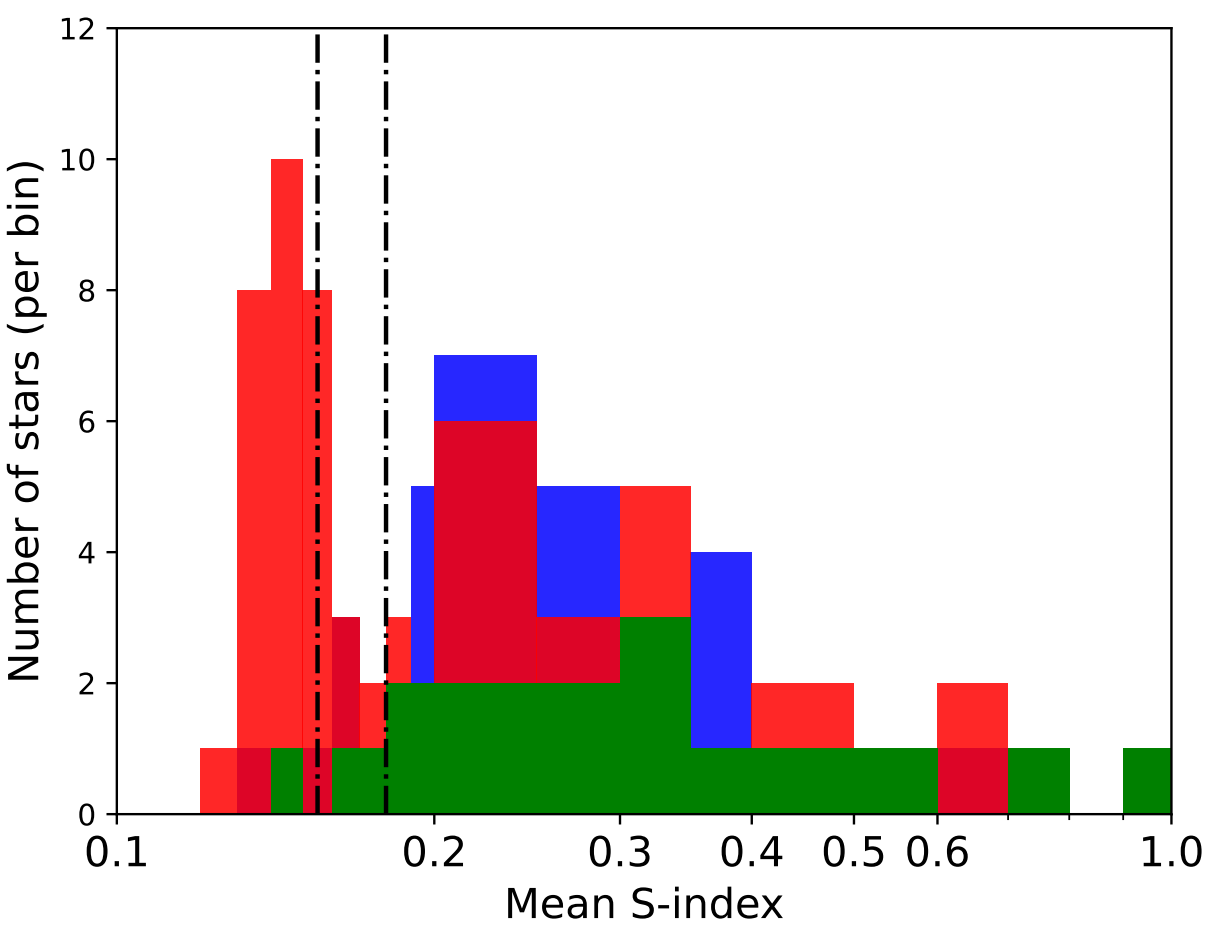

Fig. 13: Mean S-index distribution for stars without any detected cyclic variability (red histogram), those with excellent cycles (blue histogram), and the remaining sample (green histogram). The S-index range observed by TIGRE for the Sun is within the vertical dash-dotted lines (see text for details).

As discussed in Sect. 2, the Sun exhibited at least one well-documented low activity phase during the Maunder minimum era. It is therefore the only certain Maunder minimum star we know. This raises the obvious question whether other stars also enter comparable minimal activity episodes and which stars are the most likely candidates to do so. As demonstrated in Fig. 13, the hitherto observed solar activity is already at the lower end of the observed range of stellar activity in comparison to stars with stable cycles, in particular during the past two decades. This suggests that similar inactive phases and apparent instabilities may also occur in cyclic stars with similar strengths.

Given these circumstances it is thus reasonable to take the Sun as a reference point. In our data we find no star that demonstrably entered from a cyclic phase to a Maunder minimum phase or vice versa, yet we can address the question how stable the low-activity stars shown in Fig. 13 really are by considering the relative dispersion of the Mount Wilson/TIGRE data points (i.e., the standard deviation divided by the mean) as a function of the mean S-index. In Fig. 14 we plot the relative percentage dispersion of the Mount Wilson/TIGRE data points versus the mean observed S-index for all of our sample stars. It is apparent from Fig. 14 that the relative dispersion statistically increases with activity as measured through the mean S-index and that stars with excellent cycles (cyan data points) are found also in states of high activity. Focusing then on the low-activity portion of Fig. 14, i.e., stars with a mean S-index below the mean solar value, we may expect to find possible stellar analogs to the Maunder minimum Sun there.

Since no generally accepted and precise definition of a Maunder minimum candidate star exists, we define such a star by requiring, first, that its activity is lower than that of the Sun (a known Maunder minimum star) during the time period covered by TIGRE; that is, its S-index is below the mean solar S-index of 0.166. Note that this mean solar S-value is already lower than in previous decades Baliunas et al. (1995); Egeland et al. (2017). Second, we require the relative (percentage) dispersion of the Mount Wilson/TIGRE data to be below 3%. These admittedly arbitrarily introduced boundaries are shown as dashed lines in Fig. 14. For convenience we also consider less active stars with dispersions greater than 3% and, for discussion purposes, we split low variability stars into the "very low" activity category with mean S-indices below 0.14 (red data points in lower left box, group I) and the "low" activity category with a mean S-index in the range $0.14 < S < 0.16$ (magenta data points in the right box, group II). For convenience we also introduce a group III of stars with a mean S-index below the solar value, but with a relative (percentage) dispersion above 3%. Clearly, all these values are somewhat arbitrary, and there are other stars, which might also be Maunder minimum candidates but barely missed our boxes plotted in Fig. 14. A particularly promising candidate in this context is HD 10700 (= GJ71), represented by the low green downward data point just outside the solar level. According to Schmitt & Liefke (2004) it shows solar-like X-ray luminosities, its mean S-index is a bit over the solar level, and its long-term S-index curve possibly shows an 11 yr variability but below our significance thresholds; therefore, we do not consider HD 10700 any further. We finally note that all the relevant data for the stars in group I, II and III as well as the high activity stars with excellent cycles are listed in Table 1. In Table 1 are sorted into various groups, separated by horizontal lines and corresponding to the color scheme used in Fig. 14 according to their B-V color in three groups, which are defined by the activity properties of the respective group: the first group (until the first horizontal line) are the stars with excellent cycles, the second group are Maunder minimum candidate stars with a mean S-index below 0.14, and a third group with mean S-indices in the range $0.14 < S < 0.16$.

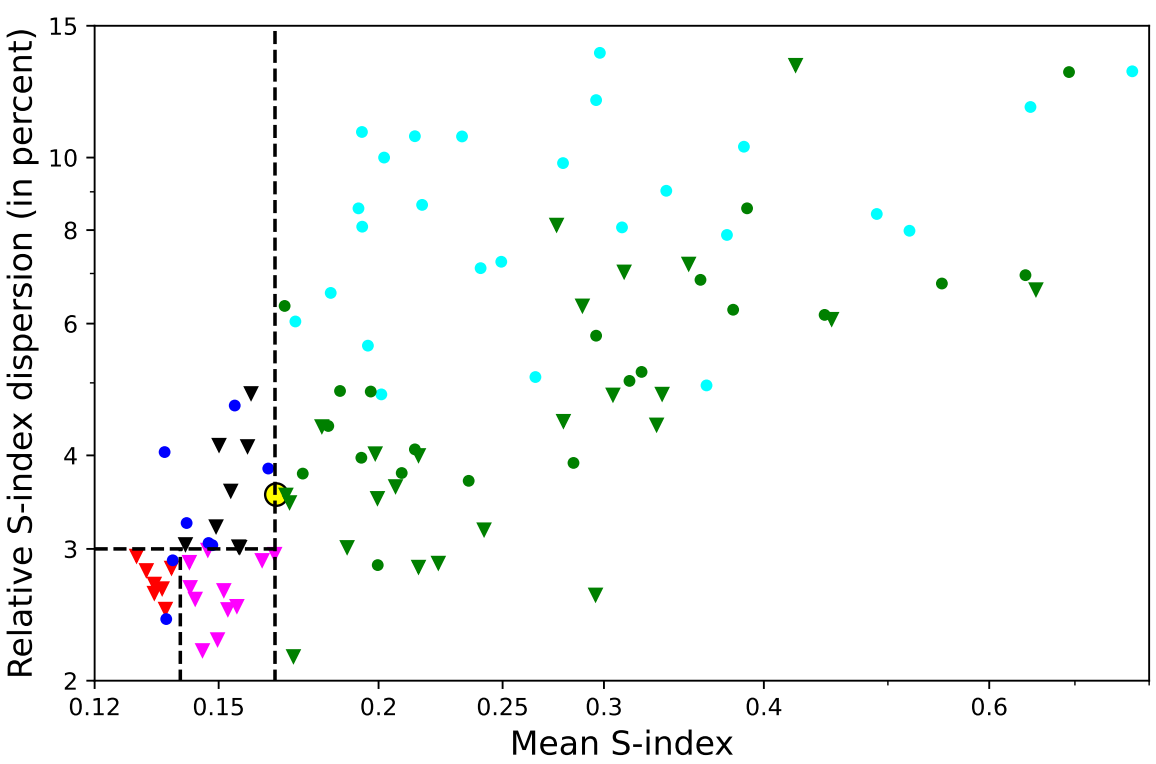

Fig. 14: Relative dispersion vs. mean S-index for all sample stars. The circles denote stars with significant cycles, and the downward triangles represent stars without measured cycles. The green and cyan symbols denote higher activity stars, and the cyan data points refer to stars with excellent cycles. The blue, black, red, and magenta data points refer to stars with lower activity; the blue data points refer to stars with cycles; and the other data points refer to stars without cycles but with different activity and variability properties. The Sun is shown as a yellow circle. The low activity stars are divided into three groups. Group I contains stars with a mean S-index below 0.14 and a relative percentage dispersion below 3. Group II contains stars with an S-index below 0.16 (but above 0.14) and a relative percentage dispersion below 3. Group III contains stars with a mean S-index below 0.16 but a relative percentage dispersion above 3 (see text for details).

It is especially interesting to examine the cyclic properties of the low-activity stars (i.e., mean S-index below solar value) and high-activity stars (i.e., mean S-index above solar value). In the low-activity category, i.e., the sum of our groups I, II, and III, we encounter 36 stars, eight of which have a cycle (four "E," one "G," and three "P"), while the rest of the sample (26) has no detectable cyclic activity: with 20 stars (i.e., the groups I and II) also having a relative variance below 3%. In contrast, in the high-activity sample (72 objects) we encounter 27 objects with excellent cycles, 21 objects with non-excellent cycles and 24 objects without any significant cyclic variability. Obviously, the fraction of stars with cycles is very different for stars with mean S-indices above solar levels (63% if all cycles are considered, 38 % if only excellent cycles are considered), while for stars below solar levels we find fractions of 24% and 12%, respectively. While clearly the formal errors on these numbers are substantial because of small number statistics, it is also clear that the fraction of cyclic stars is much lower for stars below solar activity levels.

Our specific interest in the context of stellar activity then specifically aims at the stellar Maunder minimum candidate stars as shown in Fig. 14 because they are likely to show us the distant future of solar activity. In the "very low activity" box (group I) there are nine stars satisfying our criteria with their S-index light curves shown in Fig. F.5. For some of the stars there is also SSS coverage (for HD 9562 and HD 13421), which partially fills the gap between the Mount Wilson and TIGRE observations. In the second "low activity" box (group II) there is a further group of 11 stars, which also show activity values below the Sun and the S-index light curves of which are shown in Fig. F.6. X-ray emission is naturally a good way to check and verify whether these stars can truly be considered Maunder minimum candidates. Unfortunately, only for the stars HD 13421, HD 22072, HD 29645, HD 43587, HD 89744, HD107213, HD 124570, HD 158614, HD 207978, and HD 216385, i.e., ten out of 19 objects, have sufficiently sensitive X-ray observations available; a detailed presentation of the available X-ray data on these objects is given in Appendix B. Moreover, with the possible exception of HD 187013, all stars show evidence for very low X-ray surface fluxes, thereby supporting their Maunder minimum candidacies.

### 7.2. *Summary of activity properties of Maunder minimum candidate stars*

From our long-term S-index monitoring we can identify a group of stars that have low S-indices and low relative dispersion in their temporal S-index behavior. The concept of a so-called "basal flux" of stellar chromospheres was introduced by Schrijver (1987), when he realized that even when the photospheric contributions to the S-index were removed, a residual chromospheric contribution remained, i.e., the basal flux. In other words, the state of bottom activity of a cool star is not one without any chromosphere, but rather one with a chromosphere at the basal flux; a more recent discussion of basal fluxes is given by Mittag et al. (2013), who present formulae to compute basal fluxes for stars with given effective temperature and luminosity class. Application of these formulae shows that our two classes of Maunder minimum candidate stars have S-indices very close to the "expected" basal fluxes.

Consequently one expects that a low activity state is also attained at other wavelength ranges, for example at X-rays. Our detailed investigation of the X-ray properties of our Maunder minimum candidate stars, which have been the subject of sensitive X-ray observations, shows that with the exception of the star HD 187013 all objects have low detected X-ray luminosities leading to low mean X-ray surface fluxes. HD 187013 exhibits, if it is a Maunder minimum candidate star, somewhat uncomfortable X-ray properties. In particular, if indeed, the detection as a *XMM-Newton* slew survey source is real and not due to optical contamination, one must conclude that HD 187013 can produce substantial amounts of X-ray emission temporarily. It is somewhat odd that the closer companion to HD 187013 remained undetected as an X-ray source (cf. Fig. B.2) and clearly a more sensitive X-ray observation of the HD 187013 is called for in particular with respect to the X-ray spectrum of this source.

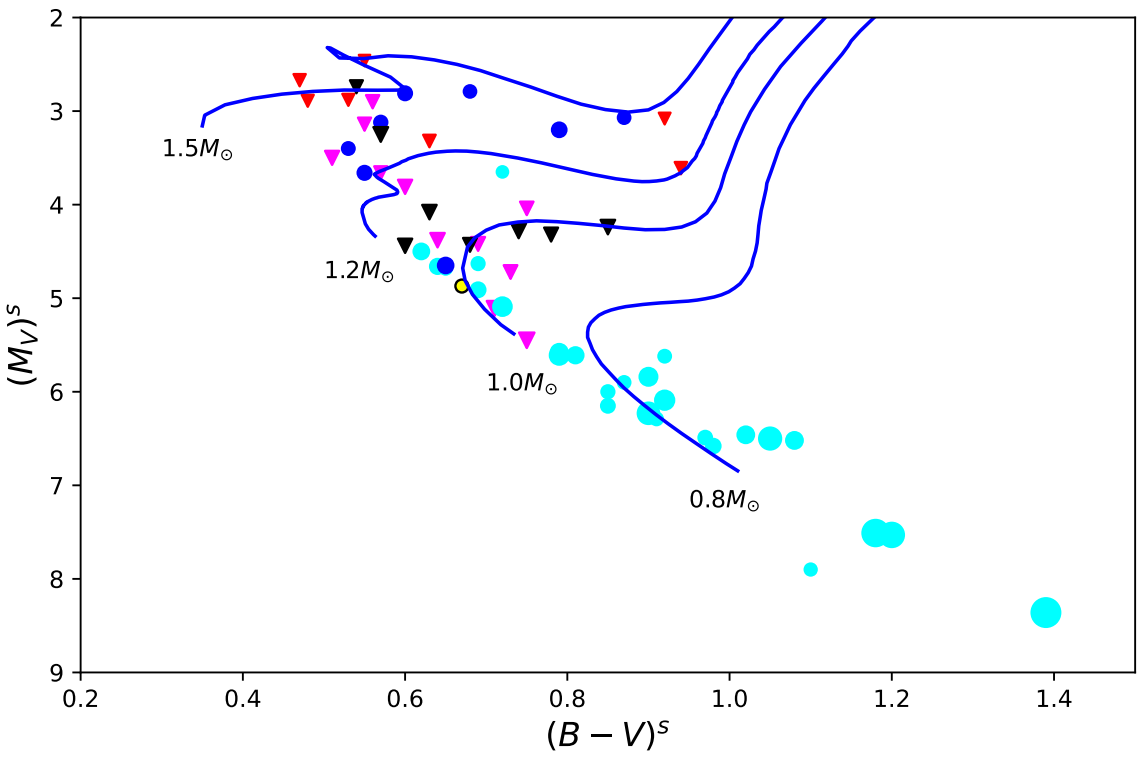

Fig. 15: HR diagram of sample stars with low activity and sample stars with high activity with excellent cycles, without cycles, and with the same color coding as in Fig. 14. The masses for the evolution tracks are indicated (see text for details).

## 8. HRD placement and clues to activity coevolution

### 8.1. General considerations and analysis procedures

To empirically understand the dependence of solar-type activity on stellar mass and evolution, it is of significant interest to examine the positions of cyclic and noncyclic stars in the Hertzsprung-Russell diagram and to compare them to Maunder minimum candidate type stars, since these states of activity represent the two activity forms we know from our Sun. To obtain a clearer picture and avoid possible ambiguities we focus here on the one hand on the low activity stars with mean S-indices below the mean solar value, and, on the other hand, on high activity stars with excellent cycles and those without any cycles, leaving out high activity stars with less significant cycles. Our analysis is based on the data summarized in Table 1. Since evolutionary tracks are usually transformed to the Johnson color system, we used the Johnson system with B-V colors and $M_V$ magnitudes, as provided by the SIMBAD database. The *Gaia* DR3 parallaxes were employed to compute accurate distances and absolute magnitudes for our sample stars. We also required the metallicity [Fe/H], for which we used [Fe/H]-values as listed in the Copenhagen-Geneva catalog (Holmberg et al. 2009) whenever available. We also list the TIGRE cycle period whenever available and the mean S-index derived from the TIGRE data.

Table 1: Sample stars

| Star | B-V | $M_v$ | [Fe/H] | [B-V]' | $M_v$' | P/yrs | Type | $< S >$ |
|---|---|---|---|---|---|---|---|---|
| **Sun** | 0.66 | 4.83 | -0.04 | 0.68 | 4.87 | 11.2 | E | 0.166 |
| Group I | | | | | | | | |
| HD107213 | 0.50 | 2.93 | 0.48 | 2.89 | +0.09 | n/a | N | 0.134 |
| HD89744 | 0.54 | 2.79 | 0.47 | 2.67 | +0.18 | n/a | N | 0.136 |
| HD124570 | 0.54 | 2.9 | 0.53 | 2.88 | +0.07 | n/a | N | 0.134 |
| HD13421 | 0.56 | 2.5 | 0.55 | 2.46 | +0.11 | n/a | N | 0.129 |
| HD9562 | 0.64 | 3.39 | 0.63 | 3.32 | +0.13 | n/a | N | 0.138 |
| HD22072$^{+}$) | 0.89 | 3.05 | 0.92 | 3.08 | n/a | n/a | N | 0.132 |
| HD23249$^{+}$) | 0.92 | 3.58 | 0.94 | 3.61 | n/a | n/a | N | 0.136 |
| Group II | | | | | | | | |
| HD207978$^{**}$) | 0.42 | 3.37 | 0.6 | 3.81 | -0.51 | n/a | N | 0.153 |
| HD187013 | 0.47 | 3.38 | 0.51 | 3.5 | -0.11 | n/a | N | 0.151 |
| HD159332 | 0.48 | 2.71 | 0.56 | 2.9 | -0.19 | n/a | N | 0.144 |
| HD216385 | 0.48 | 2.98 | 0.55 | 3.14 | -0.17 | n/a | N | 0.143 |
| HD45067 | 0.56 | 3.56 | 0.57 | 3.66 | -0.08 | n/a | N | 0.142 |
| HD142373$^{**}$) | 0.56 | 3.61 | 0.75 | 4.04 | -0.50 | n/a | N | 0.146 |
| HD143761 | 0.60 | 4.19 | 0.69 | 4.42 | -0.25 | n/a | N | 0.150 |
| HD43587 | 0.61 | 4.26 | 0.64 | 4.38 | -0.11 | n/a | N | 0.155 |
| HD126053 | 0.63 | 5.06 | 0.75 | 5.45 | -0.39 | n/a | N | 0.166 |
| HD115617 | 0.71 | 5.08 | 0.71 | 5.11 | 0.00 | n/a | N | 0.162 |
| HD190360$^{+}$) | 0.73 | 4.69 | 0.73 | 4.72 | n/a | n/a | N | 0.147 |
| Group III | | | | | | | | |
| HD212754 | 0.52 | 2.79 | 0.54 | 2.74 | +0.10 | n/a | N | 0.141 |
| HD141004 | 0.60 | 4.04 | 0.63 | 4.08 | -0.01 | n/a | N | 0.156 |
| HD12235 | 0.62 | 3.37 | 0.57 | 3.25 | +0.18 | n/a | N | 0.158 |
| HD217014 | 0.67 | 4.5 | 0.68 | 4.43 | +0.12 | n/a | N | 0.150 |
| HD3795 | 0.70 | 3.86 | 0.85 | 4.24 | -0.42 | n/a | N | 0.156 |
| HD178428 | 0.70 | 4.45 | 0.6 | 4.44 | +0.05 | n/a | N | 0.153 |
| HD158614 | 0.72 | 4.24 | 0.74 | 4.28 | -0.01 | n/a | N | 0.160 |
| HD182572 | 0.77 | 4.29 | 0.78 | 4.32 | n/a | n/a | N | 0.150 |
| HD76572 | 0.43 | 2.53 | 0.6 | 2.81 | -0.34 | 16.4 | E | 0.147 |
| HD136202 | 0.54 | 3.06 | 0.57 | 3.12 | -0.04 | 15.07 | P | 0.142 |
| HD187691 | 0.55 | 3.66 | 0.55 | 3.66 | n/a | 8.66 | G | 0.148 |
| HD29645 | 0.57 | 3.48 | 0.53 | 3.4 | +0.14 | 14.95 | P | 0.138 |
| HD100180$^{*}$) | 0.57 | 4.5 | 0.65 | 4.65 | -0.15 | 3.71 | E | 0.16 |
| HD161239$^{+}$) | 0.65 | 2.75 | 0.68 | 2.79 | n/a | 4.96 | E | 0.136 |
| HD219834A$^{+}$) | 0.80 | 3.16 | 0.79 | 3.2 | n/a | 17.28 | E | 0.154 |
| HD188512 | 0.86 | 3.04 | 0.87 | 3.07 | n/a | 4.69 | P | 0.137 |
| HD114710 | 0.57 | 4.43 | 0.69 | 4.63 | -0.21 | 17.52 | E | 0.201 |
| HD154417 | 0.57 | 4.44 | 0.62 | 4.5 | -0.08 | 9.57 | E | 0.265 |
| HD78366 | 0.60 | 4.55 | 0.64 | 4.66 | -0.10 | 12.21 | E | 0.249 |
| HD190406 | 0.61 | 4.53 | 0.65 | 4.68 | -0.12 | 16.35 | E | 0.196 |
| HD72905 | 0.62 | 4.84 | 0.72 | 5.09 | -0.25 | 11.9 | E | 0.361 |
| HD81809$^{*}$) | 0.64 | 3.35 | 0.72 | 3.65 | -0.29 | 8.13 | E | 0.172 |
| HD76151 | 0.67 | 4.87 | 0.69 | 4.91 | -0.04 | 19.18 | E | 0.240 |
| HD101501 | 0.72 | 5.43 | 0.79 | 5.58 | -0.12 | 18.46 | E | 0.310 |
| HD103095$^{**}$) | 0.75 | 6.64 | 1.1 | 7.9 | -1.36 | 7.09 | E | 0.184 |
| HD152391 | 0.76 | 5.51 | 0.79 | 5.61 | -0.08 | 8.95 | E | 0.374 |
| HD185144 | 0.80 | 5.88 | 0.85 | 6.15 | -0.24 | 6.28 | E | 0.216 |
| HD10780 | 0.81 | 5.62 | 0.81 | 5.61 | +0.05 | 9.69 | E | 0.279 |
| HD26965 | 0.82 | 5.93 | 0.85 | 6.0 | -0.04 | 9.95 | E | 0.202 |
| HD149661 | 0.82 | 5.79 | 0.9 | 5.84 | -0.01 | 11.76 | E | 0.336 |
| HD10476 | 0.84 | 5.83 | 0.87 | 5.9 | -0.04 | 10.1 | E | 0.193 |
| HD165341A | 0.86 | 5.77 | 0.92 | 6.09 | -0.29 | 5.03 | E | 0.386 |
| HD166620 | 0.87 | 6.17 | 0.91 | 6.29 | -0.08 | 16.3 | E | 0.194 |
| HD4628 | 0.88 | 6.38 | 0.98 | 6.58 | -0.17 | 8.95 | E | 0.232 |
| HD22049 | 0.88 | 6.19 | 0.9 | 6.23 | 0.00 | 10.77 | E | 0.490 |
| HD219834B$^{+}$) | 0.91 | 5.58 | 0.92 | 5.62 | n/a | 8.92 | E | 0.194 |
| HD115404$^{+}$) | 0.93 | 6.46 | 1.05 | 6.5 | n/a | 10.87 | E | 0.520 |
| HD160346 | 0.96 | 6.32 | 1.02 | 6.46 | -0.09 | 7.05 | E | 0.296 |
| HD16160 | 0.98 | 6.53 | 0.97 | 6.49 | +0.08 | 13.01 | E | 0.214 |
| HD32147$^{+}$) | 1.06 | 6.48 | 1.08 | 6.52 | n/a | 10.68 | E | 0.298 |
| HD156026$^{+}$) | 1.16 | 7.47 | 1.18 | 7.51 | n/a | 14.83 | E | 0.776 |
| HD201091$^{+}$) | 1.18 | 7.49 | 1.2 | 7.53 | n/a | 7.19 | E | 0.646 |
| HD201092$^{+}$) | 1.37 | 8.31 | 1.39 | 8.36 | n/a | 11.2 | E | 0.958 |

$^{*}$) Primary only, close fainter companion considered $^{**}$) Very metal-poor star, large shift is uncertain $^{+}$) Assume solar metallicity

**Notes.** List of stars used for Figs. 14 and 15), divided into three groups as shown in Fig. 14. The table provides the star name, B-V colors from SIMBAD, distance moduli $M_v$ from DR3 parallaxes, shifted B-$V^s$, and $M_v{}^s$ values to project the HRD positions of sample stars with the listed metallicities onto a common reference set of evolution tracks with Z=0.02, (see text for details and Fig. 14), cycle periods as derived from the combined data TIGRE and Mount Wilson data, the derived cycle type, and the mean S-indices for the period 2013-2024 (see text for details).

## 9. HRD placement and clues to activity coevolution

### 9.1. General considerations and analysis procedures

To empirically understand the dependence of solar-type activity on stellar mass and evolution, it is of significant interest to examine the positions of cyclic and noncyclic stars in the Hertzsprung-Russell diagram and to compare them to Maunder minimum candidate type stars, since these states of activity represent the two activity forms we know from our Sun. To obtain a clearer picture and avoid possible ambiguities we focus here on the one hand on the low activity stars with mean S-indices below the mean solar value, and, on the other hand, on high activity stars with excellent cycles and those without any cycles, leaving out high activity stars with less significant cycles. Our analysis is based on the data summarized in Table 1. Since evolutionary tracks are usually transformed to the Johnson color system, we used the Johnson system with B-V colors and $M_V$ magnitudes, as provided by the SIMBAD database. The *Gaia* DR3 parallaxes were employed to compute accurate distances and absolute magnitudes for our sample stars. We also required the metallicity [Fe/H], for which we used [Fe/H]-values as listed in the Copenhagen-Geneva catalog (Holmberg et al. 2009) whenever available. We also list the TIGRE cycle period whenever available and the mean S-index derived from the TIGRE data.

### 9.2. A direct comparison of all stars: compensating different metallicities

We now proceed to consider the evolutionary status of our Maunder minimum candidate stars by placing them in the HRD. Color and luminosity of a star, and consequently the placement of an evolutionary track in the HRD depend on metallicity in addition to the stellar mass. Since our sample stars have rather different metallicities (cf. Table 1), they cannot be compared directly in an HRD with respect to their evolution and mass. To achieve this end and to project stars of different metallicities onto a common reference set of evolutionary tracks, we follow the approach detailed by Schröder et al. (2013) by applying shifts in B-V and $M_v$ to project each star from its individual metallicity, with respect to evolutionary tracks with a metallicity of Z=0.02. The applied shifts of the individual star's placements then result in shifted values $[B\text{-}V]^s$ and $M_v{}^s$, which are also listed in Table 1).

As demonstrated by Schröder et al. (2013), comparing evolutionary tracks computed with the well-tested Eggleton code with abundances differing by a factor of 2, there is little change in these shifts along the main sequence; for example, a metallicity difference of [Fe/H]=-0.3 results in a displacement compensation of +0.09 in B-V with +0.29 in $M_v$ for 0.8 $M_\odot$, +0.08 (B-V) with +0.25 ($M_v$) for 1.2 $M_\odot$, and +0.12 (B-V) with +0.23 ($M_v$) for 1.5 $M_\odot$. In this procedure we also assume that the Sun, compared to Z=0.02 abundances, already has a small metallicity difference of [Fe/H]=-0.04, which adds to the Copenhagen-Geneva [Fe/H] values of the stars listed there. If we did not apply these shifts in the HRD, metal-poor stars would look more massive and younger than they really are, because by virtue of their low metallicity they are hotter and more luminous than their more metal-rich siblings of otherwise the same mass and evolutionary advance.

In Fig. 15 we show an HR diagram for our sample stars with high activity and "excellent" cycles (cyan circles) as well as our three categories of noncyclic low activity stars (downward triangles) and cyclic low-activity stars (blue circles), constructed from the above described shifted B-V color and shifted absolute magnitudes. The Sun is indicated by a yellow circle. For better comparability the color coding in Fig. 15 is the same as that in Fig. 14. We also show evolutionary tracks (computed for Z = 0.02) for masses of 1.5, 1.2, 1.0, and 0.8 solar masses to enable a rough mass assessment for the sample stars. The Sun, the only unequivocal Maunder minimum star, is also shown in Fig. 15, accompanied in its HRD vicinity by a few other cyclic stars as well as Maunder minimum candidate stars, and finally we note that the symbol size for the various data points in Fig. 15 is an indicator for the magnitude of the underlying S-index value. Figure 15 demonstrates that our group I stars are somewhat evolved more massive stars, while our group II stars are somewhat less massive. The group III stars lie somewhat in between, and half the stars in group III do show cyclic activity, half of them (i.e., 4 objects) even with excellent cycles.

### 9.3. The grand picture: coevolution and mass dependence

The HRD shown in Fig. 15 represents one of the basic results of our study, and we summarize the following four rather straightforward findings. First, active stars with excellent cycles (i.e., the data points represented by the cyan filled circles in Fig. 15) cover almost the whole mass range on the main sequence for stars with convective envelopes, i.e., 0.7 to 1.5 solar masses, even though they appear less common among F-stars. Second, among the lower-mass stars on the lower luminosity main sequence, where the convective envelopes reach deeper and magnetic braking timescales are longer, we find much stronger activity among the cyclic stars (as for other activity forms as well, such as multi-periodic or chaotic). Third, cyclic stars with a low solar-level-like activity (i.e., the blue data points in Fig. 15 ) are found not only close to the main sequence but also further away; here, we note that among the faster evolving, more massive F-stars, cyclic cases are close to the turn-off, while the majority of highly active, cyclic solar-type stars seem closer to the zero-age main sequence. The overlap seen on the lower main sequence between strong and modest cyclic activity is naturally explained by the fact that early stellar evolution runs in parallel to the ZAMS for stars below one solar mass. And fourth, as to the far end of magnetic activity coevolution, the Maunder-Minimum candidate stars (as identified by the two boxes in the lower left corner of Fig. 14) are all in the second half of their main sequence life or already nearing its end. The least active stars in terms of their S-indices (in the left box in Fig. 14) are more massive than their solar-type Maunder minimum candidate cousins in the right box; these stars do not lie near the zero-age main sequence in contrast to the very active stars, but turn out to be evolved main-sequence stars.

This finding is in line with those by Schröder et al. (2013), who showed that Maunder minimum candidate type stars, fading to a remnant magnetic activity of little variation, are on average more evolved than cyclic stars. Furthermore, in Appendix B we show that the X-ray activity of those stars (for which sufficiently sensitive data are available) is also low; therefore, these stars are of special interest to understanding the distant activity future of the Sun and the characteristics of its remnant activity.

During the recent solar Maunder minimum active regions were essentially absent most of the time. Remnant emissions in such a state must therefore be indicative of local processes. The stars identified by our work to represent such a low activity state are mainly Maunder minimum candidate star in group II of Fig. 14, they also emit X-rays near the lower limit of coronal emission in the HRD and the inactive solar corona (cf. Schmitt 1997; Schröder et al. 2012), coincidental with reaching an S-index value formed under the presence of only the basal chromospheric flux (ca. 0.140...0.155) outside active regions. It is clear that such an inactive, "basal" chromosphere has little room for much variation.

Remarkably, the occurrence of such inactive phases and apparent instability of the solar cycle indicates that solar activity is at the lower end of the range (as measured by the S-index), when compared to the Mount Wilson stars with stable cycles. In particular, the past two decades have placed the Sun there with an average S-value during the TIGRE observing period of only 0.166. But even in more active phases, for example in the 1950s, when in maximum years S-values reached 0.19 – 0.20 (somewhat dependent on the applied calibration), the solar cycle still maintained a rather modest strength, compared to other cyclic stars similar to the Sun.

In this context the study of Usoskin et al. (2007) is particularly relevant, who found that the Sun spent around 17% of the time over the last 11300 yrs in a grand minimum state. If we were to observe the Sun 18 times (the number of our Maunder minimum candidate stars) for 60 yrs (the timescale covered by our joint Mount Wilson-TIGRE data) at random epochs selected within the last 11300 yrs, a Monte Carlo simulation lets us expect to observe $6 \pm 2$ Maunder minimum entry or exit events, yet we observed none since our TIGRE data does not support the entry of HD166620 into a Maunder minimum phase as argued by Baum et al. (2022). Therefore, it appears that our group of Maunder minimum candidate stars spends considerably more than 17% of the time in their minimum states.

## 10. Conclusions

The extension of the timeline of chromospheric activity monitoring to more than five decades through an additional 11 years of calibrated data obtained with TIGRE, helped us find additional cycles of longer periods and better define the significance and quality of their periodicities. Clearly, longer and possibly denser sampling may reveal even more cyclic variations. About a third of the sample stars have very significant ("excellent") cycles and, as demonstrated by our paper, we were able to roughly *double* the stars with known cycles within the same Mount Wilson stellar sample of main-sequence stars. Newly confirmed cases of cyclic activity predominantly come from stars with higher activity than observed for the Sun, with longer and shorter cycle periods. Among F-type stars we find cycles shorter than one year. However, stars with lower activity than the Sun are far less likely to exhibit cyclic variability.

On the low mass end of the stellar sample, where the convective envelopes become increasingly massive, we find the strongest chromospheric emission levels in all forms of magnetic activity (cyclic, chaotic, multi-period), consistent with existing knowledge. Around one solar mass, not surprisingly, the typical cycle periods are on the order of 10 years. The most stable, best-defined ("excellent") cycles comprise many solar-like stars, which are more active and less evolved than our host-star. Yet, a number of stars with uncertain cycle properties still remain, which require studies on even longer timescales. Among the more active stars, one finds definite variations, but it is not clear whether these variations are solely stochastic or not.

In addition, we identified a group of eighteen stars, which show – over the sampled time range of $\approx 60$ years – mean S-indices below 0.166, i.e., below the mean S-index of the Sun observed by TIGRE in the past two decades, and a relative dispersion of less than 3% again below the observed solar variability level. We refer to these stars as Maunder minimum candidate stars, and for about half the sample we can demonstrate their very low activity level also at X-ray wavelengths. Of course, we have no way of knowing whether these stars remain permanently in their presently observed low-activity states or whether they eventually turn a state with cyclically variable activity as the present-day Sun, which according to Usoskin et al. (2007) spends around 17% of the time in its Maunder minimum state. If these solar grand minima timescales are also typical for our sample, we would expect to observe about half a dozen of such changes from cyclic activity to a Maunder minimum state or vice versa. Yet, somewhat surprisingly we observed none.

Our results suggest that main-sequence stars that have evolved beyond the solar state enter a phase of noncyclic remnant activity. In the case of fast evolving F-stars, they may even reach a state of almost zero activity. In such cases, their chromospheric emission fades into the basal flux level, and coronal X-ray emission also approaches a lower limit, particularly in our group I of Maunder Minimum candidate stars identified in Fig. 14. Their G-type inactive equivalents (group II of Fig. 14) may permanently be in a state comparable to that of the inactive Sun in its occasional episodes during its grand minima. Yet, such transitions must be relatively rare and, therefore, require far larger sample sizes and possibly more extended monitoring for a meaningful study.

In terms of the emerging empirical picture of magnetic activity coevolution with stellar structure, mixed with some star-to-star variation, we observe a clear decline of the chromospheric activity strength with the evolutionary advance on the main sequence in accordance with previous authors (Schröder et al. 2013; Oláh et al. 2016): early main-sequence stars are very active, varying in a chaotic or multi-periodic manner, or have only poor cycles before they enter a stable cyclic stage. Initially, such younger cyclic stars have a much stronger activity than the present-day Sun. Some of these stars make good proxies for the young Sun and the irradiation of the young Earth with UV, far-UV, EUV, and X-ray radiation. More evolved stars (comparable to the Sun or a bit beyond on the main sequence), show a similarly modest cyclic activity as the present-day Sun and the currently observed solar mean S-index fits very well to the lower end of the chromospheric activity range of cyclic main-sequence stars.

In consequence, the outlook – on a timescale of a few billion years – for our Sun is then more and longer phases of such inactivity than the present 17% as estimated by Usoskin et al. (2007). We further expect the solar cycle to become increasingly unstable, and solar activity ought to eventually fade to permanently very low levels as encountered in the Maunder minimum candidate stars of our sample. While the Sun appears to have entered and exited from its Maunder minimum stage 27 times over the last 9300 yrs, i.e., on average every 340 yrs, it remains a challenge to identity stellar analogs of such events. This should be possible with sufficiently large and carefully selected stellar samples and sufficiently dense observing cadences.

*Acknowledgements.* We thank our anonymous referee for the careful scrutiny of our paper and the many suggestions which improved our paper considerably. The HK_Project_v1995_NSO data derive from the Mount Wilson Observatory HK Project, which was supported by both public and private funds through the Carnegie Observatories, the Mount Wilson Institute, and the Harvard-Smithsonian Center for Astrophysics starting in 1966 and continuing for over 36 years. These data are the result of the dedicated work of O. Wilson, A. Vaughan, G. Preston, D. Duncan, S. Baliunas, and many others. The TIGRE project was only made possible by the continuous support of the Universities of Hamburg and Guanajuato, and a significant contribution by the University of Liège. We also want to thank our collaborator and friend Prof. Dr. Rauw for the long partnership within the TIGRE project and specifically for his detailed comments on our paper. In addition we acknowledge support, mainly for travel and collaboration, by a larger number of DFG and Conacyt projects.

## Appendix A: Some notes on individual stars

With TIGRE we were not able to obtain S-index measurements for all of the 111 sample stars, for the closer visual binaries we encountered occasional technical problems in the data acquisition, and we discuss those objects in the following section.

### A.1. 70 Oph

The star 70 Oph (HD 165341A + HD 165341B) is a nearby binary system consisting of two dwarf stars of spectral type K0 and KIV, discovered already by Herschel 250 years ago; its orbit is well determined Strand (1937) with a period of 87.85 yr and a semimajor axis of 4.558 arcsec. During the times of the TIGRE observations the system was near maximum separation of about 6.4 arcsec, yet the TIGRE acquisition system was programmed to lock on the brightest star in the field of view, i.e., HD 165341A. Therefore, we did not plan but only inadvertently took TIGRE data for HD 165341B. While the Mount Wilson database contains separate light curves for both stars, the light curve of the fainter star is, however, much noisier. The S-indices of the two components are quite different and a few of our S-index measurements of HD 165341A must clearly refer to HD 165341B, and these values were then removed from the TIGRE data on HD 165341A "by hand."

### A.2. $\eta$ CrB

The star $\eta$ CrB (= HD 137107 + HD 137108) is a visual, but close binary consisting of two stars of spectral type G2V also detected by Herschel more than 200 years ago. The orbit determination by Breakiron et al. (1975) yields a binary period of 41.65 yr and a semimajor axis of 0.9 arcsec; thus, the two binary components cannot be separated in neither the Mount Wilson nor the TIGRE time series.

### A.3. $\xi$ Boo

$\xi$ Boo (= HD 131156AB) is a well known BY Dra type binary consisting of the components G7Ve and K5Ve. Söderhjelm (1999) determined an orbital period of 151 yr and a semimajor axis of 4.94 arcsec. Wood & Linsky (2010) present high-resolution X-ray observations obtained with the LETGS on board the *Chandra* Observatory, which clearly separate the two binary components at an apparent angular distance of 6 arcsec; somewhat surprisingly the A component accounts for almost 90% of the system's X-ray luminosity. In contrast, Baliunas et al. (1995) find the S-indices of 0.76 and 1.17 for the A and B components, respectively. As far as TIGRE is concerned, the TIGRE light curves refer – as in the case of 70 Oph – to the brighter A component of the system, and we have no TIGRE data for HD 131156B.

### A.4. 94 Aqr

The star 94 Aqr (= HD 219834AB) is actually a triple system. The A component is itself a visual binary (with sub-arc second separation; cf. Docobo et al. 2018), while the secondary is about 13 arcsec away. Docobo et al. (2018) carried out a detailed investigation of spectroscopic and photometric observations of this system and conclude that the Aa and Ab components of the system are a G8 subgiant and a K2-K3 dwarf with an orbital period of 6.3 yr; the more distant B component at an angular distance of 12.4 arcsec is of spectral type K2. The TIGRE observations refer to the A component.

### A.5. 36 Oph

The 36 Oph system is actually a visual triple system; however, the third component (= Gaia DR3 4109034455276324608) is at an angular distance of 732 arcsec away from the central binary 36 Oph (= HD 155885 + HD155886), which in turn consists of two K dwarfs at an angular separation of about 5 arcsec; the orbital period of 36 Oph is somewhat uncertain but must be a few hundred years Irwin et al. (1996). *Gaia* photometry suggests that the magnitudes and colors of these two components are extremely similar. Wood & Linsky (2006) present *Chandra* observations of 36 Oph, which show two clearly separated sources at a distance of 5.1 arcsec. The X-ray luminosities of the two components are very similar, with the A component being slightly brighter. Baliunas et al. (1995) present two separate light curves for 36 Oph, yet the S-indices of the two components are rather similar, while with TIGRE we only have one light curve.

### A.6. 61 Cyg

The star 61 Cyg (HD 201091 + HD 201092) is a well known wide binary system consisting of two K dwarfs with an orbit period of almost 700 years. The apparent binary separation is more than 20 arcsec; therefore, the system can be easily separated. TIGRE took almost 200 observations of this system. Unfortunately, occasionally the TIGRE acquisition system acquired the "wrong" star. Fortunately, the S-indices of the two components are very different, so that the spectral data can be correctly attributed "by hand" to the respective counterparts.

### A.7. V891 Tau + V774 Tau

The case of V891 Tau (= HD26913 = HR1321) and V774 Tau (= HD26923 = HR1322) is similar to the case of 61 Cyg. These two stars are of spectral type G0IVe (HR1322) and G8Ve (HR1321) and form a proper motion pair located at the same physical distance of 22 pc with an angular separation of 64 arcsec. TIGRE took almost 200 spectra of the system components and, again, occasionally the "wrong" star was acquired. Both stars are quite active and fortunately the S-indices of the fainter component are larger than those of the brighter component so that the data can again be correctly attributed "by hand."

## Appendix B: X-ray properties of Maunder minimum candidate stars

The Maunder minimum candidate stars shown in the two boxes in Fig. 14 are obviously located near the bottom of the observed H and K activity range; thus, it behooves us well to check whether these stars are also at the bottom of the observed activity range at other wavelengths (e.g., X-rays). We therefore checked the corresponding archives from the ROSAT, *XMM-Newton* and *Chandra* observatories for available observations of these sources, which we briefly discuss in the following. Stars not discussed in the following have not been observed at X-ray wavelengths in dedicated observations and have remained undetected in the ROSAT all-sky survey, implying that the thus obtained upper limits are not that sensitive to explicitly demonstrate the Maunder minimum character of these sources; these remarks refer to the stars HD 13421, HD 22072, HD 29645, HD 43587,

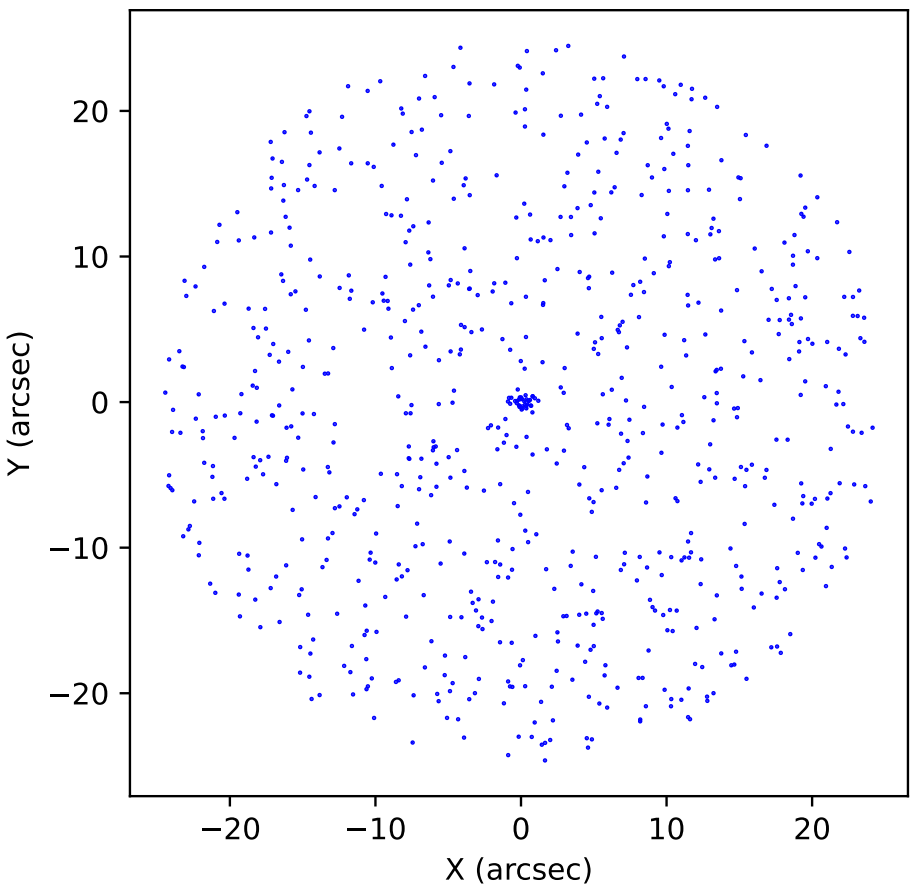

Fig. B.1: Central portion of the *Chandra* Observatory HRC-I image of HD-143761. All registered events are referenced with respect to the nominal FK5 position of HD-143761.

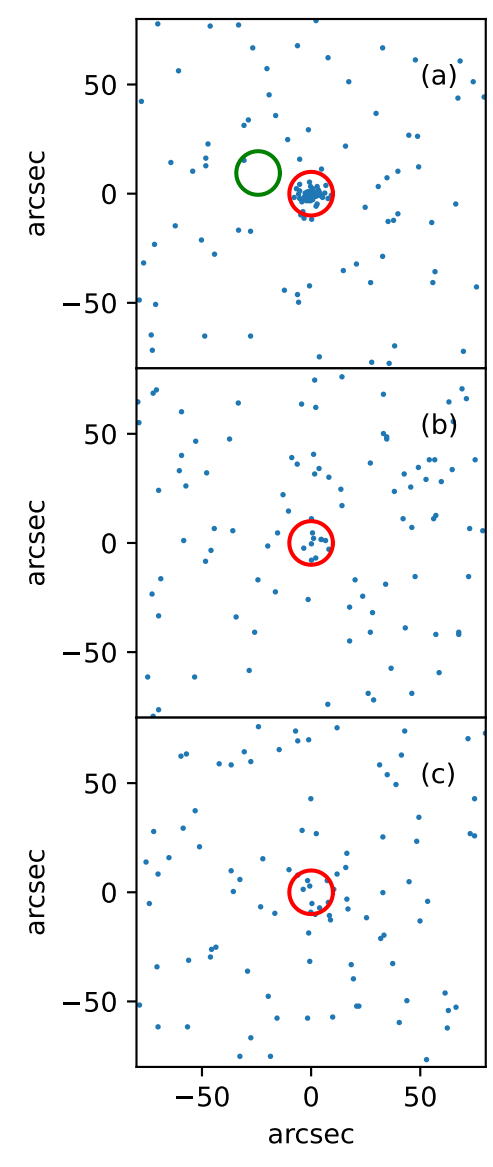

Fig. B.2: ROSAT HRI image of HD 187013 system. Panels (a), (b), and (c) refer to the different components of the HD 187013 (= 17 Cyg) system, described in detail in the text.

HD 89744, HD107213, HD 124570, HD 158614, HD 207978 and HD 216385. In the following we discuss the X-ray properties of the remaining stars and in particular investigate whether their measured X-ray surface fluxes are near the minimum mean surface flux of ≈ $10^4$ erg/cm$^2$/s identified by Schmitt (1997) in a sample of nearby solar-like stars.

### B.1. HD 9562

HD 9562 was observed with the ROSAT PSPC, the results are described by Hempelmann et al. (1996). At the position of HD 9562 a weak soft X-ray source was detected with an observed count rate of 0.0035 cts/s. translates into an X-ray luminosity of course ≈ 5.4 $10^{26}$ erg/s, which is indeed very low for a G-type star. Using the results for the stellar radius derived by Gaia Collaboration et al. (2018), we compute a mean X-ray surface flux of ≈ 2500 erg/cm$^2$/s, which again is very low.

### B.2. HD 23249

HD 23249 was observed with the ROSAT PSPC, the results are described by Schmitt & Liefke (2004). HD 23249 was detected both in the ROSAT all-sky survey (RASS) and the pointing program, with X-ray luminosities of log $L_X$ = 27.05 and 27.25, respectively. According to the results of Gaia Collaboration et al. (2018), HD 23249 has a rather large radius of 2.93 $R_\odot$ albeit with considerable uncertainty. Using the nominal values we find a mean X-ray surface flux of 2400 erg/cm$^2$/s.

### B.3. HD 142373

Schmitt & Liefke (2004) report X-ray detections of HD 142373 both with the ROSAT PSPC and HRI yielding X-ray luminosities of log $L_X$ = 26.95 and 26.92, respectively, consistent with the X-ray luminosity of a low-activity late F-type star. Using the radius derived by Gaia Collaboration et al. (2018), we find an X-ray surface flux of ≈ 5400 erg/cm$^2$/s.

### B.4. HD 143761

Considerable confusion exists concerning the X-ray properties of HD 143761. While one finds in SIMBAD the source 1E 1559.1+3327 associated with HD 143761, a comparison of the coordinates does reveal that this X-ray source cannot be positionally identified with HD 143761; also, HD 143761 remained undetected in the ROSAT all-sky survey (Hempelmann et al. 1996). In the *XMM-Newton* archive one finds a source right at the position of HD 143761, yet an inspection of the source's X-ray properties shows that this signal is very likely produced by optical contamination and not X-rays. Binder et al. (2024) report the non-detection of HD 143761 with the ACIS-instrument on board the *Chandra* Observatory, however, they fail to mention that an observation of HD 143761 with the HRC-I on board the *Chandra* Observatory also exists. We downloaded these data from the *Chandra* Observatory archive (ObsId 22308, sequence number 201278) and display the central portion of the image centered on the nominal FK5 coordinates of HD 143761 in Fig. B.1. Without the necessity of any statistics, a source is clearly present at the position of HD 143761, for which we derive a count rate of 2.7 $10^{-3}$ ct/s with a formal error of 20%. With the *Chandra* Observatory HRC-I little to no energy information is available for the registered counting events. Anticipating a low-activity, low temperature source similar, for example, to 51 Peg, we refer to the discussion by Poppenhäger et al. (2009) and use an energy conversion factor of 9.4 $10^{-12}$ erg/cm$^2$/ct to derive an X-ray flux of 2.5 $10^{-14}$ erg/cm$^2$/s, which translates into an X-ray luminosity of 9.2 $10^{26}$ erg/s using a distance of 17.5 pc. With a radius of R = 1.33 $R_\odot$ as derived by *Gaia* one computes a mean X-ray surface flux of ≈ 8500 erg/cm$^2$/s for HD 143761, which puts HD 143761 at the bottom of the observed X-ray surface flux scale (cf. Schmitt 1997) and suggests it to be a "coronal hole" star.

### B.5. HD 187013

The case of HD 187013 is also a little more complicated. According to Hempelmann et al. (1996) and Schmitt & Liefke (2004) the star HD 187013 was detected in the ROSAT all-sky survey with an X-ray luminosity of log $L_X$ = 28.62, a slightly higher X-ray luminosity of log $L_X$ = 29.06 was derived from *XMM-Newton* slew survey observations, i.e., at rather high X-ray luminosities.

The star HD 187013 ( = 17 Cyg) is actually a member of a multiple stellar system, consisting of five visual stars altogether. In the *Gaia* DR3 data release one finds 5 entries (Gaia DR3 2035035223783462400, Gaia DR3 2035035258143206016, Gaia DR3 2035020964491479552, Gaia DR3 2035021028896869888, Gaia DR3 2035038900276000000), which are located at more or less the same distance and more or less share the same proper motion with the possible exception of the last entry. The first four entries have essentially the same distance of 21 pc, while the fifth entry is located a bit further away at 25.5 pc. The first object is the by far brightest one and of spectral type F; this star is the target of the CaII H and K monitoring. All other stars are of later type, the second to fourth entry are K-type stars. The first and second star and the third and fourth object form close pairs with angular distances of 26 arcsec and 3 arcsec, respectively, while the angular distance between the two pairs is more than 700 arcsec. The fifth object, Gaia DR3 2035038900276000000, is very red with a Gaia color BP-RP = 3.47, corresponding to a very late M-type star at an angular distance of again more than 700 arcsec.

Fortunately a pointing with ROSAT HRI has been obtained (sequence number 201289) that covers all of the components of the 17 Cyg system. The relevant image portions are displayed in Fig. B.2. In Fig. B.2 panel (a) refers to the bright component Gaia DR3 2035035223783462400 and the common proper motion star Gaia DR3 2035035258143206016, panel (a) refers to the closer visual pair of K-type stars (i.e., Gaia DR3 2035021028896869888 and Gaia DR3 2035038900276000000) and panel (c) to the late-type M star Gaia DR3 2035020964491479552. Figure B.2 (in panel a) shows a clear detection of HD 187013, while its visual companion Gaia DR3 2035035258143206016 remains undetected. The circles displayed in Fig. B.2 have a radius of 10 arcsec and should contain almost all of the photons from a source at their respective centers. Somewhat surprisingly we also detect weak signals from the K binary proper motion component and the late M-component. In the respective detection cell, we obtain ten and seven counts, respectively, while 1.1 counts are expected from background. Note that in the ”official” source lists, these sources do not appear, yet we searched for sources only at their known relative (with respect to HD 183017) positions. Numerical experiments using a detection cell with a radius of 10 arcsec placed at random positions on the HRI detector show that finding seven and – even more so – ten counts is very unlikely with probabilities below $10^{-5}$. We therefore believe that these detections are actually real. Yet, needless to say, these detections must be treated with some caution and require confirmation with deeper pointings. In particular, if the detection of the very red object Gaia DR3 2035020964491479552 is indeed real, it would be truly unusual. From the ROSAT HRI observations we derive a somewhat lower (as compared to the PSPC) X-ray luminosity of $9 \times 10^{27}$ erg/s, which translates into a mean X-ray surface flux of 57000 erg/cm$^2$/s.

#### B.5.1. HD 190360

The star HD 190360 (= GJ 777 A, Gaia DR3 2029433521248546304) is also member of a binary system. Its companion (GJ 777 B, Gaia DR3 2029432043779954432) is an M dwarf with virtually identical distance and 3D space motion, but located at an angular distance of 178 arcsec with respect to GJ 777 A. GJ 777 A was observed twice with the *XMM-Newton* observatory. The stacked 4XMM catalog contains the source 4XMM J200326.8+295158, right at the expected position of GJ 777 B at epoch 2005 with a flux of 8.6 $10^{-14}$ erg/cm$^2$/s, which corresponds to an X-ray luminosity of 2.6 $10^{27}$ erg/s at the distance of GJ 777 B. However, there is no cataloged source at the position of GJ 777A. The XMM-Newton images suggest that a very weak source might be present at the position of GJ 777 A; however, that source is certainly not statistically significant. We therefore adopted a flux upper limit of ≈ 3 $10^{-14}$ erg/cm$^2$/s for GJ 777A, which corresponds to an upper limit of $9 \times 10^{26}$ erg/s for its X-ray luminosity $L_X$ and ≈ 7250 erg/cm$^2$/s as an upper limit for the mean X-ray surface flux. Thus, it is clear that GJ 777 A is in all likelihood also one of the ”coronal hole” stars and a deeper X-ray observation would certainly be worthwhile.

## Appendix C: Yearly breakup of TIGRE observations

Table C.1 provides a summary of the available TIGRE observations of the Mount Wilson stars presented by Baliunas et al. (1995): we specifically provide the star names and the number of observations carried out in each of the years 2013 - 2024: for each star we quote the HD number, the number of observations per year, and in the last column a flag indicating whether SSS data are also available; see text for details.

Table C.1: Yearly break down of TIGRE observations.

| HD number | 2013 | 2014 | 2015 | 2016 | 2017 | 2018 | 2019 | 2020 | 2021 | 2022 | 2023 | 2024 | Total | SSS |
|---|---|---|---|---|---|---|---|---|---|---|---|---|---|---|
| HD 1835 | 0 | 6 | 6 | 0 | 7 | 3 | 7 | 1 | 2 | 2 | 7 | 0 | 41 | Y |
| HD 2454 | 10 | 9 | 8 | 0 | 16 | 20 | 10 | 3 | 2 | 4 | 7 | 0 | 89 | |
| HD 3229 | 7 | 8 | 6 | 0 | 17 | 19 | 10 | 3 | 8 | 3 | 8 | 0 | 89 | |
| HD 3443 | 6 | 7 | 7 | 0 | 7 | 1 | 8 | 1 | 2 | 2 | 7 | 0 | 48 | |
| HD 3651 | 7 | 8 | 8 | 0 | 5 | 1 | 6 | 1 | 2 | 1 | 7 | 0 | 46 | |
| HD 3795 | 8 | 7 | 7 | 0 | 6 | 1 | 7 | 2 | 2 | 4 | 8 | 0 | 52 | |
| HD 4628 | 7 | 7 | 6 | 0 | 6 | 2 | 7 | 1 | 2 | 4 | 7 | 0 | 49 | |
| HD 6920 | 19 | 16 | 8 | 0 | 6 | 2 | 7 | 12 | 8 | 7 | 9 | 0 | 94 | Y |
| HD 9562 | 0 | 6 | 7 | 0 | 6 | 2 | 6 | 2 | 2 | 3 | 5 | 1 | 40 | Y |
| HD 10476 | 16 | 12 | 11 | 0 | 14 | 2 | 6 | 3 | 4 | 2 | 4 | 1 | 75 | Y |
| HD 10700 | 13 | 17 | 17 | 1 | 19 | 4 | 25 | 10 | 8 | 7 | 8 | 0 | 129 | Y |
| HD 10780 | 0 | 6 | 35 | 8 | 28 | 31 | 47 | 35 | 40 | 45 | 42 | 3 | 320 | |
| HD 12235 | 8 | 9 | 7 | 1 | 14 | 2 | 7 | 2 | 2 | 2 | 4 | 1 | 59 | |
| HD 13421 | 17 | 14 | 17 | 0 | 15 | 13 | 23 | 10 | 7 | 2 | 5 | 0 | 123 | Y |
| HD 16160 | 10 | 10 | 7 | 1 | 17 | 3 | 5 | 12 | 5 | 2 | 4 | 0 | 76 | |
| HD 16673 | 17 | 14 | 10 | 0 | 12 | 18 | 28 | 45 | 19 | 22 | 24 | 3 | 212 | |
| HD 17925 | 18 | 33 | 15 | 0 | 13 | 5 | 10 | 33 | 7 | 6 | 6 | 0 | 146 | |
| HD 18256 | 9 | 10 | 5 | 1 | 9 | 11 | 7 | 2 | 5 | 2 | 15 | 4 | 80 | Y |
| HD 20630 | 0 | 30 | 19 | 4 | 15 | 2 | 52 | 50 | 31 | 46 | 62 | 4 | 315 | Y |
| HD 22049 | 15 | 25 | 20 | 5 | 12 | 7 | 112 | 59 | 33 | 51 | 60 | 4 | 403 | |
| HD 22072 | 19 | 17 | 8 | 3 | 5 | 1 | 4 | 7 | 8 | 8 | 8 | 0 | 88 | |
| HD 23249 | 20 | 20 | 12 | 4 | 10 | 5 | 25 | 26 | 15 | 12 | 9 | 1 | 159 | |
| HD 25998 | 22 | 49 | 15 | 1 | 10 | 13 | 23 | 19 | 12 | 17 | 22 | 5 | 208 | |
| HD 26913 | 9 | 10 | 11 | 1 | 10 | 3 | 8 | 26 | 6 | 4 | 6 | 0 | 94 | |
| HD 26923 | 19 | 38 | 12 | 1 | 11 | 10 | 23 | 30 | 5 | 14 | 19 | 4 | 186 | |
| HD 26965 | 8 | 8 | 6 | 1 | 11 | 6 | 10 | 23 | 7 | 6 | 8 | 1 | 95 | |
| HD 29645 | 14 | 9 | 10 | 1 | 4 | 14 | 7 | 3 | 3 | 1 | 4 | 0 | 70 | |
| HD 30495 | 21 | 44 | 30 | 9 | 12 | 12 | 11 | 25 | 11 | 13 | 9 | 2 | 199 | Y |
| HD 32147 | 8 | 9 | 5 | 1 | 4 | 4 | 8 | 5 | 3 | 2 | 6 | 1 | 56 | Y |
| HD 33608 | 12 | 10 | 8 | 1 | 9 | 15 | 9 | 7 | 3 | 2 | 5 | 1 | 82 | |
| HD 35296 | 20 | 49 | 29 | 5 | 17 | 25 | 21 | 14 | 20 | 12 | 9 | 2 | 223 | Y |
| HD 37394 | 22 | 33 | 25 | 8 | 11 | 41 | 12 | 8 | 6 | 10 | 7 | 3 | 186 | |
| HD 39587 | 16 | 37 | 19 | 30 | 15 | 30 | 31 | 20 | 23 | 16 | 24 | 10 | 271 | Y |
| HD 43587 | 10 | 17 | 16 | 13 | 17 | 18 | 74 | 35 | 19 | 44 | 62 | 26 | 351 | Y |
| HD 45067 | 16 | 24 | 11 | 2 | 11 | 17 | 23 | 12 | 11 | 15 | 20 | 8 | 170 | |
| HD 61421 | 17 | 46 | 47 | 12 | 27 | 27 | 18 | 17 | 35 | 10 | 10 | 9 | 275 | |
| HD 72905 | 16 | 52 | 36 | 14 | 34 | 29 | 11 | 13 | 10 | 32 | 8 | 6 | 261 | Y |
| HD 75332 | 16 | 40 | 38 | 6 | 13 | 23 | 27 | 20 | 16 | 20 | 28 | 19 | 266 | Y |
| HD 76151 | 0 | 4 | 12 | 10 | 15 | 6 | 54 | 40 | 51 | 52 | 59 | 30 | 333 | Y |
| HD 76572 | 8 | 12 | 9 | 3 | 11 | 17 | 6 | 3 | 0 | 0 | 7 | 5 | 81 | |
| HD 78366 | 8 | 13 | 9 | 4 | 5 | 20 | 11 | 8 | 5 | 3 | 12 | 4 | 102 | Y |
| HD 81809 | 0 | 3 | 11 | 7 | 14 | 6 | 9 | 9 | 19 | 12 | 9 | 8 | 107 | Y |
| HD 82443 | 8 | 12 | 14 | 1 | 20 | 10 | 9 | 12 | 22 | 12 | 9 | 9 | 138 | |
| HD 82885 | 0 | 3 | 11 | 33 | 50 | 40 | 32 | 39 | 34 | 39 | 48 | 53 | 382 | |
| HD 88355 | 6 | 11 | 6 | 5 | 14 | 19 | 10 | 3 | 0 | 3 | 9 | 4 | 90 | |
| HD 88737 | 4 | 11 | 8 | 4 | 12 | 16 | 6 | 2 | 10 | 1 | 8 | 5 | 87 | |
| HD 89744 | 15 | 30 | 13 | 4 | 16 | 19 | 8 | 8 | 9 | 7 | 7 | 6 | 142 | |
| HD 95735 | 6 | 16 | 16 | 0 | 8 | 11 | 16 | 5 | 4 | 2 | 3 | 15 | 102 | |
| HD 97334 | 0 | 32 | 14 | 21 | 28 | 55 | 9 | 14 | 11 | 26 | 9 | 9 | 228 | Y |
| HD 100180 | 6 | 9 | 8 | 5 | 19 | 24 | 13 | 8 | 13 | 24 | 9 | 9 | 147 | |
| HD 100563 | 7 | 27 | 8 | 5 | 13 | 22 | 27 | 21 | 34 | 18 | 35 | 17 | 234 | |
| HD 101501 | 8 | 30 | 27 | 16 | 21 | 47 | 25 | 54 | 47 | 83 | 65 | 44 | 467 | |
| HD 103095 | 5 | 9 | 9 | 9 | 8 | 5 | 8 | 4 | 0 | 9 | 8 | 8 | 82 | |
| HD 106516 | 6 | 26 | 12 | 7 | 14 | 26 | 14 | 16 | 33 | 10 | 9 | 9 | 182 | |
| HD 107213 | 4 | 12 | 9 | 5 | 14 | 27 | 8 | 3 | 0 | 2 | 10 | 4 | 98 | |
| HD 111456 | 3 | 12 | 9 | 5 | 18 | 26 | 10 | 7 | 3 | 10 | 7 | 8 | 118 | |
| HD 114378 | 4 | 27 | 14 | 5 | 18 | 23 | 26 | 23 | 17 | 47 | 34 | 16 | 254 | |
| HD 114710 | 3 | 20 | 25 | 18 | 25 | 42 | 48 | 38 | 24 | 65 | 75 | 57 | 440 | |
| HD 115043 | 6 | 37 | 27 | 27 | 50 | 51 | 83 | 58 | 29 | 52 | 62 | 29 | 511 | Y |
| HD 115383 | 4 | 27 | 13 | 5 | 34 | 26 | 31 | 27 | 19 | 40 | 29 | 21 | 276 | Y |
| HD 115404 | 1 | 13 | 10 | 5 | 28 | 9 | 9 | 9 | 4 | 16 | 12 | 6 | 122 | |
| HD 115617 | 4 | 35 | 25 | 5 | 32 | 9 | 11 | 14 | 12 | 11 | 9 | 10 | 177 | Y |
| HD 120136 | 5 | 82 | 72 | 89 | 165 | 114 | 152 | 112 | 78 | 109 | 152 | 117 | 1247 | Y |
| HD 124570 | 4 | 29 | 14 | 7 | 20 | 25 | 10 | 12 | 7 | 32 | 7 | 8 | 175 | |
| HD 124850 | 4 | 32 | 15 | 7 | 19 | 27 | 16 | 14 | 18 | 7 | 10 | 9 | 178 | |
| HD 126053 | 6 | 41 | 37 | 24 | 37 | 9 | 62 | 40 | 31 | 46 | 70 | 42 | 445 | Y |
| HD 129333 | 6 | 50 | 12 | 0 | 9 | 12 | 8 | 64 | 13 | 9 | 8 | 10 | 201 | |
| HD 131156A | 0 | 0 | 14 | 6 | 36 | 6 | 8 | 12 | 8 | 32 | 7 | 8 | 137 | |
| HD 131156B | 0 | 0 | 0 | 0 | 0 | 0 | 0 | 0 | 0 | 0 | 0 | 0 | 0 | |
| HD 136202 | 0 | 10 | 7 | 4 | 24 | 24 | 7 | 5 | 2 | 4 | 11 | 5 | 103 | |
| HD 137107 | 4 | 27 | 14 | 4 | 31 | 23 | 31 | 30 | 19 | 30 | 60 | 24 | 297 | |
| HD 141004 | 0 | 16 | 25 | 14 | 31 | 56 | 32 | 17 | 6 | 37 | 85 | 36 | 355 | Y |
| HD 142373 | 4 | 35 | 12 | 5 | 25 | 24 | 14 | 13 | 7 | 7 | 7 | 8 | 161 | |
| HD 143761 | 0 | 0 | 21 | 28 | 39 | 6 | 58 | 43 | 30 | 49 | 63 | 41 | 378 | Y |
| HD 149661 | 1 | 8 | 8 | 4 | 9 | 6 | 10 | 4 | 1 | 9 | 7 | 7 | 74 | Y |
| HD 152391 | 1 | 8 | 8 | 3 | 37 | 7 | 10 | 3 | 1 | 3 | 9 | 5 | 95 | Y |
| HD 154417 | 1 | 8 | 6 | 3 | 8 | 23 | 13 | 4 | 1 | 3 | 9 | 4 | 83 | Y |
| HD 155885 | 0 | 0 | 4 | 3 | 3 | 11 | 8 | 3 | 3 | 6 | 7 | 8 | 56 | |
| HD 155886 | 0 | 0 | 0 | 0 | 0 | 0 | 0 | 0 | 0 | 0 | 0 | 0 | 0 | |
| HD 156026 | 1 | 9 | 3 | 1 | 9 | 9 | 12 | 10 | 9 | 6 | 9 | 5 | 83 | |
| HD 157856 | 1 | 7 | 5 | 3 | 26 | 25 | 11 | 4 | 1 | 3 | 9 | 4 | 99 | |
| HD 158614 | 1 | 13 | 11 | 3 | 8 | 6 | 9 | 13 | 9 | 9 | 7 | 8 | 97 | |
| HD 159332 | 1 | 15 | 7 | 3 | 21 | 22 | 9 | 3 | 0 | 8 | 7 | 7 | 103 | |
| HD 160346 | 1 | 6 | 3 | 4 | 9 | 6 | 9 | 2 | 0 | 9 | 7 | 7 | 63 | Y |
| HD 161239 | 1 | 12 | 5 | 3 | 38 | 7 | 8 | 4 | 0 | 8 | 7 | 7 | 100 | |
| HD 165341A | 1 | 7 | 4 | 3 | 9 | 10 | 10 | 5 | 6 | 9 | 9 | 7 | 80 | |
| HD 165341B | 0 | 0 | 0 | 0 | 0 | 0 | 0 | 0 | 0 | 0 | 0 | 0 | 0 | |
| HD 166620 | 1 | 8 | 5 | 2 | 11 | 8 | 10 | 3 | 0 | 8 | 8 | 7 | 71 | |
| HD 176051 | 2 | 8 | 5 | 2 | 10 | 28 | 9 | 4 | 1 | 5 | 7 | 3 | 84 | |
| HD 176095 | 1 | 3 | 6 | 2 | 23 | 28 | 11 | 2 | 1 | 4 | 7 | 3 | 91 | |
| HD 178428 | 3 | 7 | 9 | 0 | 15 | 13 | 20 | 4 | 1 | 5 | 35 | 8 | 120 | |
| HD 182101 | 2 | 3 | 6 | 2 | 22 | 24 | 7 | 5 | 1 | 5 | 7 | 3 | 87 | |
| HD 182572 | 5 | 12 | 18 | 2 | 25 | 25 | 31 | 36 | 13 | 27 | 27 | 9 | 230 | |
| HD 185144 | 3 | 2 | 7 | 2 | 13 | 11 | 16 | 11 | 4 | 9 | 7 | 7 | 92 | Y |
| HD 187013 | 4 | 7 | 6 | 2 | 32 | 28 | 12 | 4 | 1 | 7 | 7 | 7 | 117 | |
| HD 187691 | 3 | 4 | 7 | 2 | 26 | 25 | 10 | 3 | 1 | 5 | 7 | 3 | 96 | Y |
| HD 188512 | 3 | 8 | 9 | 2 | 10 | 9 | 13 | 6 | 1 | 7 | 8 | 6 | 82 | |
| HD 190007 | 3 | 0 | 1 | 0 | 7 | 5 | 6 | 3 | 0 | 1 | 8 | 3 | 37 | |
| HD 190360 | 7 | 5 | 4 | 2 | 10 | 5 | 9 | 10 | 7 | 7 | 7 | 6 | 79 | |
| HD 190406 | 2 | 2 | 4 | 1 | 9 | 4 | 8 | 3 | 1 | 7 | 7 | 5 | 53 | |
| HD 194012 | 5 | 5 | 6 | 1 | 28 | 24 | 12 | 5 | 0 | 4 | 7 | 3 | 100 | |
| HD 201091 | 9 | 18 | 11 | 4 | 19 | 5 | 13 | 12 | 8 | 24 | 7 | 5 | 135 | Y |
| HD 201092 | 0 | 3 | 2 | 4 | 11 | 4 | 11 | 3 | 0 | 3 | 7 | 2 | 50 | Y |
| HD 206860 | 5 | 6 | 7 | 0 | 8 | 19 | 13 | 7 | 1 | 6 | 7 | 3 | 82 | Y |
| HD 207978 | 11 | 21 | 14 | 1 | 26 | 24 | 10 | 10 | 8 | 9 | 7 | 4 | 145 | |
| HD 212754 | 6 | 7 | 5 | 0 | 21 | 22 | 10 | 5 | 2 | 5 | 7 | 2 | 92 | |
| HD 216385 | 12 | 15 | 8 | 0 | 23 | 23 | 11 | 10 | 7 | 8 | 7 | 5 | 129 | |
| HD 217014 | 12 | 15 | 20 | 0 | 33 | 9 | 31 | 15 | 9 | 40 | 60 | 16 | 260 | Y |
| HD 219834A | 7 | 8 | 6 | 0 | 10 | 4 | 10 | 4 | 2 | 4 | 8 | 4 | 67 | |
| HD 219834B | 0 | 0 | 0 | 0 | 5 | 2 | 9 | 1 | 1 | 4 | 7 | 0 | 29 | |
| HD 224930 | 11 | 31 | 17 | 0 | 23 | 5 | 10 | 5 | 2 | 4 | 7 | 3 | 118 | Y |

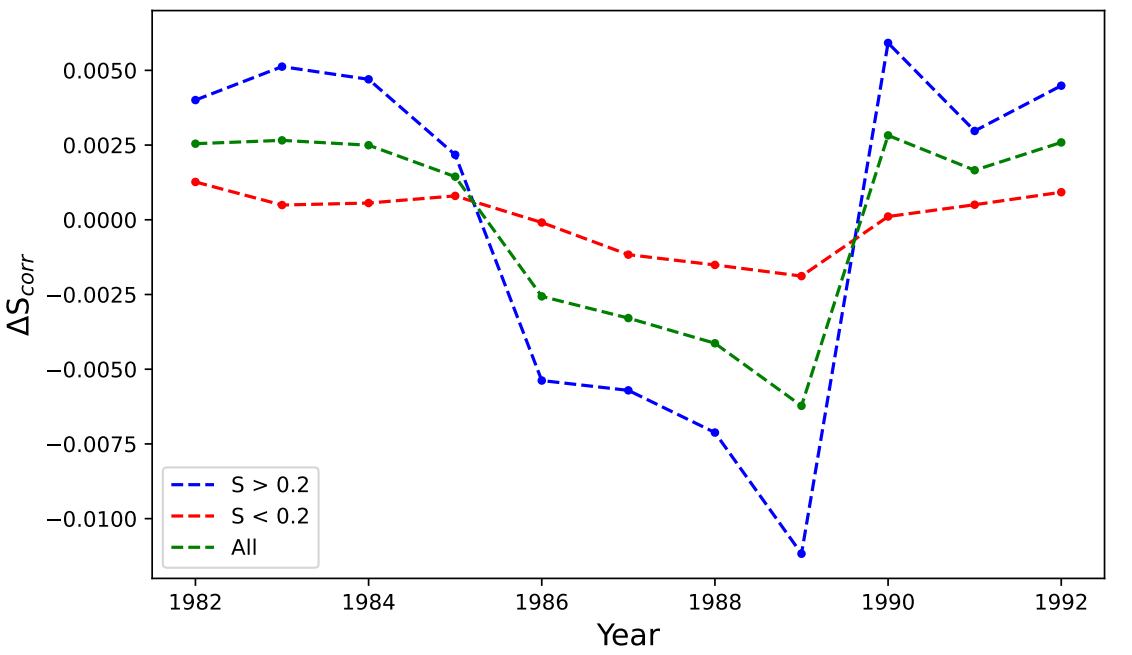

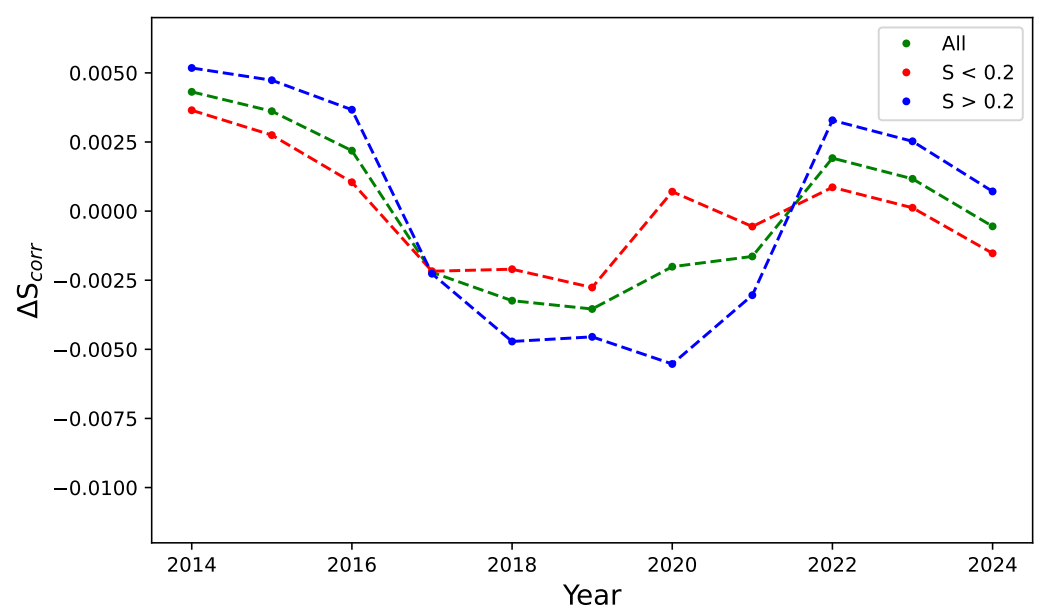

Fig. C.1: Residual S-index $S_{corr}$ variations for the Mount Wilson data (left panel) and TIGRE data (right panel). Note that both plots are on the same scale (see text for more details).

## Appendix D: calibration stability

In Sect. 3.5 we presented our investigations of the cross calibration and stability of the S-index time series used in this paper. As shown by Fig. 2, the mean S-index of a stellar sample depends - obviously - on the composition of the individual sample members. To obtain the clearest results, one should therefore use time series containing the same stars. To this end we identified in our TIGRE data those stars with available data for every year, vice versa, in the case of the Mount Wilson S-index data we selected the years 1982 - 1992, i.e., the same time lapse as the TIGRE data, and again identified those stars with available data for every year in this time frame. For both datasets we formed yearly averages of the measured S-indices, considering three groups of stars, i.e., low activity stars with a mean S-index below 0.2, high activity stars with a mean S-index above 0.2, and all stars. For each of these groups we constructed the summed yearly S-index light curves plotted in Fig. C.1; for better comparability of the different groups, the respective mean summed S-indices have been subtracted thus leaving the residual $S_{corr}$.

Both the $S_corr$-time graphs for the Mount Wilson and TIGRE data look quite similar. The runs of the variations with time observed in the different groups are similar, yet the most active stars show, not surprisingly, the largest variations, while the low activity stars show smaller variations in their residual S-indices $S_{corr}$. For the Mount Wilson data the maximally encountered variations amount to $S_{corr} \approx 0.017$, for the low activity stars we find $S_{corr} \approx 0.003$ for the TIGRE data the corresponding numbers are 0.011 and 0.006, respectively; we thus conclude that the resulting $S_{corr}$-values are of similar size. We have no firm explanation for these variations, and they could simply be the result of not fully averaged out natural sample variations, yet these variations might also result from using supposedly inactive calibration stars, which may show small residual variations as well.

## Appendix E: Cycle analysis results

### E.1. Stars with detected cycles

In Table E.1 we provide a list of those 58 stars with significantly detected cycles according to our classification scheme outlined in Sect. 5.2.

Table E.1: Stars with significantly detected cycles.

| Star name | Per (yr) | log(-P) | $Per_{sec}$ (yr) | $Per_{third}$ (yr) | $S_{obs}$ | $\sigma_{obs}$ | $S_{mod}$ | $\sigma_{mod}$ | ΔAmp | Δ $\phi$ | T | MW | O2016 |
|---|---|---|---|---|---|---|---|---|---|---|---|---|---|
| Sun | 11.27 | 202.62 | n.a. | n.a. | 0.1664 | 0.0059 | 0.1659 | 0.0042 | 0.0118 | 0.4288 | E | 10.0E | 11S |
| HD3651 | 12.55 | 4.37 | n.a. | n.a. | 0.1689 | 0.0107 | 0.0094 | 0.1682 | 0.0157 | 0.4810 | G | 13.8G | 11.6S |
| HD4628 | 8.95 | 58.73 | n.a. | n.a. | 0.2324 | 0.0248 | 0.0142 | 0.2287 | 0.0559 | 0.4770 | E | 8.37E | 8.94S |
| HD6920 | 15.87 | 3.35 | 12.45 P | n.a. | 0.1939 | 0.0077 | 0.0068 | 0.1949 | 0.0111 | 0.5511 | F | n.a. | n.a. |
| HD10476 | 10.10 | 20.81 | n.a. | n.a. | 0.1929 | 0.0165 | 0.0118 | 0.1912 | 0.0332 | 0.5010 | E | 9.6E | 9.8S |
| HD10780 | 9.69 | 12.54 | 7.55 G | 13.20 G | 0.2787 | 0.0274 | 0.0202 | 0.2782 | 0.0531 | 0.5711 | E | n.a. | n.a. |
| HD16160 | 13.01 | 42.03 | n.a. | n.a. | 0.2135 | 0.0228 | 0.0132 | 0.2128 | 0.0518 | 0.4329 | E | 3.2E | 12.1S |
| HD18256 | 7.29 | 2.73 | n.a. | n.a. | 0.1827 | 0.0080 | 0.0074 | 0.1825 | 0.0086 | 0.5691 | P | 6.8F | n.a. |
| HD20630 | 5.80 | 2.36 | n.a. | n.a. | 0.3569 | 0.0245 | 0.0228 | 0.3586 | 0.0261 | 0.4289 | P | 5.6F | 5.32C |
| HD22049 | 10.77 | 21.42 | 8.22 E | 15.07 E | 0.4901 | 0.0412 | 0.0320 | 0.4942 | 0.0715 | 0.4970 | E | n.a. | n.a. |
| HD26913 | 10.34 | 3.21 | 7.89 P | n.a. | 0.3785 | 0.0237 | 0.0212 | 0.3797 | 0.0311 | 0.5150 | F | 7.8F | n.a. |
| HD26923 | 10.70 | 3.84 | 8.30 P | 5.53 P | 0.2840 | 0.0111 | 0.0100 | 0.2843 | 0.0135 | 0.5010 | F | n.a. | n.a. |
| HD26965 | 9.95 | 64.40 | 15.24 G | n.a. | 0.2020 | 0.0202 | 0.0096 | 0.2054 | 0.0491 | 0.4269 | E | 10.1E | 10.0S |
| HD29645 | 14.95 | 2.26 | n.a. | n.a. | 0.1381 | 0.0040 | 0.0038 | 0.1379 | 0.0043 | 0.4228 | P | n.a. | n.a. |
| HD30495 | 11.04 | 2.59 | n.a. | n.a. | 0.2958 | 0.0171 | 0.0153 | 0.2962 | 0.0215 | 0.4770 | P | n.a. | n.a. |
| HD32147 | 10.68 | 14.14 | n.a. | n.a. | 0.2978 | 0.0411 | 0.0288 | 0.2944 | 0.0833 | 0.4970 | E | 11.1E | 10.6S |
| HD33608 | 11.64 | 2.20 | n.a. | n.a. | 0.2135 | 0.0087 | 0.0077 | 0.2141 | 0.0106 | 0.4228 | P | n.a. | n.a. |
| HD39587 | 15.77 | 2.62 | n.a. | n.a. | 0.3141 | 0.0158 | 0.0146 | 0.3122 | 0.0180 | 0.4228 | P | n.a. | n.a. |
| HD72905 | 11.90 | 5.42 | 16.97 G | 18.87 G | 0.3608 | 0.0179 | 0.0146 | 0.3601 | 0.0282 | 0.4749 | E | n.a. | n.a. |
| HD76151 | 19.18 | 5.54 | n.a. | n.a. | 0.2403 | 0.0171 | 0.0148 | 0.2439 | 0.0263 | 0.5190 | E | 2.52F | 5.32C |
| HD76572 | 16.40 | 6.65 | 10.84 F | n.a. | 0.1473 | 0.0045 | 0.0039 | 0.1482 | 0.0077 | 0.5531 | E | 7.1P | n.a. |
| HD78366 | 12.21 | 5.98 | 15.96 E | 2.24 G | 0.2494 | 0.0181 | 0.0162 | 0.2498 | 0.0249 | 0.4689 | E | 12.2G | 13.45C |
| HD81809 | 8.13 | 48.71 | n.a. | n.a. | 0.1722 | 0.0104 | 0.0063 | 0.1716 | 0.0225 | 0.4248 | E | 8.17E | 8.69S |
| HD82443 | 10.03 | 2.45 | 7.31 P | n.a. | 0.6404 | 0.0446 | 0.0399 | 0.6419 | 0.0517 | 0.5671 | P | 2.8P | n.a. |
| HD88737 | 11.73 | 3.01 | n.a. | n.a. | 0.2352 | 0.0087 | 0.0077 | 0.2353 | 0.0108 | 0.4269 | F | 24? | n.a. |
| HD100180 | 3.71 | 5.25 | 9.95 P | n.a. | 0.1640 | 0.0063 | 0.0052 | 0.1636 | 0.0091 | 0.4449 | E | 3.56F | 13.2C |
| HD100563 | 10.36 | 2.60 | n.a. | n.a. | 0.1997 | 0.0057 | 0.0051 | 0.2000 | 0.0072 | 0.4950 | P | n.a. | n.a. |
| HD101501 | 18.46 | 5.04 | 11.25 F | 4.48 P | 0.3099 | 0.0250 | 0.0222 | 0.3141 | 0.0325 | 0.4970 | E | n.a. | n.a. |
| HD103095 | 7.09 | 45.49 | n.a. | n.a. | 0.1835 | 0.0121 | 0.0072 | 0.1854 | 0.0268 | 0.5591 | E | 7.30E | 6.95S |
| HD106516 | 5.97 | 2.38 | n.a. | n.a. | 0.2085 | 0.0079 | 0.0073 | 0.2083 | 0.0088 | 0.4329 | P | n.a. | n.a. |
| HD114710 | 17.52 | 6.34 | 5.46 P | 11.78 P | 0.2010 | 0.0097 | 0.0087 | 0.1995 | 0.0131 | 0.5050 | E | 16.6G | 16.6C |
| HD115043 | 4.26 | 3.34 | 11.93 P | 8.63 P | 0.3210 | 0.0166 | 0.0146 | 0.3214 | 0.0233 | 0.4850 | F | n.a. | n.a. |
| HD115404 | 10.87 | 5.70 | n.a. | n.a. | 0.5198 | 0.0415 | 0.0357 | 0.5201 | 0.0592 | 0.4970 | E | 2.4G | 10.8C |
| HD129333 | 12.45 | 4.06 | n.a. | n.a. | 0.5510 | 0.0374 | 0.0326 | 0.5540 | 0.0609 | 0.4689 | G | n.a. | n.a. |
| HD131156A | 11.40 | 2.99 | n.a. | n.a. | 0.4462 | 0.0275 | 0.0255 | 0.4461 | 0.0288 | 0.4930 | P | n.a. | 3.7C |
| HD136202 | 15.07 | 2.36 | n.a. | n.a. | 0.1416 | 0.0046 | 0.0043 | 0.1412 | 0.0052 | 0.4930 | P | n.a. | n.a. |
| HD137107 | 7.24 | 2.43 | n.a. | n.a. | 0.1845 | 0.0066 | 0.0059 | 0.1832 | 0.0091 | 0.5772 | P | n.a. | n.a. |
| HD149661 | 11.76 | 6.25 | 3.99 F | n.a. | 0.3356 | 0.0303 | 0.0271 | 0.3379 | 0.0399 | 0.5130 | E | 17.4G | 4.0C |
| HD152391 | 8.95 | 6.80 | 11.47 E | n.a. | 0.3743 | 0.0295 | 0.0259 | 0.3754 | 0.0412 | 0.5110 | E | 10.9E | 9.9C |
| HD154417 | 9.57 | 7.09 | 7.89 E | n.a. | 0.2652 | 0.0135 | 0.0116 | 0.2629 | 0.0197 | 0.5210 | E | 7.4F | n.a. |
| HD155885 | 17.74 | 3.23 | 9.19 P | 11.81 P | 0.3881 | 0.0332 | 0.0290 | 0.3917 | 0.0459 | 0.5651 | F | 5.7P | n.a. |
| HD156026 | 14.83 | 12.87 | 8.27 E | 10.15 E | 0.7761 | 0.1012 | 0.0809 | 0.7648 | 0.1668 | 0.5030 | E | 21.0G | 4.33C |
| HD157856 | 11.90 | 2.24 | n.a. | n.a. | 0.1972 | 0.0096 | 0.0090 | 0.1979 | 0.0098 | 0.5010 | P | 15.9F | n.a. |
| HD160346 | 7.05 | 52.21 | 4.96 P | n.a. | 0.2958 | 0.0353 | 0.0221 | 0.2949 | 0.0822 | 0.5130 | E | 7.0E | 7.35S |
| HD161239 | 4.96 | 5.98 | n.a. | n.a. | 0.1361 | 0.0055 | 0.0047 | 0.1362 | 0.0077 | 0.5772 | E | 5.7F | n.a. |
| HD165341A | 5.03 | 5.78 | 15.51 F | 7.98 P | 0.3859 | 0.0399 | 0.0326 | 0.3921 | 0.0716 | 0.5691 | E | 5.1F | 11.4C |
| HD166620 | 16.30 | 52.24 | n.a. | n.a. | 0.1942 | 0.0157 | 0.0087 | 0.1935 | 0.0380 | 0.5251 | E | 15.8E | 13.6S |
| HD176051 | 11.20 | 3.97 | 16.95 P | n.a. | 0.1745 | 0.0066 | 0.0058 | 0.1748 | 0.0103 | 0.4409 | F | 10? | n.a. |
| HD185144 | 6.28 | 26.23 | 7.98 E | 18.85 E | 0.2163 | 0.0187 | 0.0117 | 0.2113 | 0.0414 | 0.5170 | E | n.a. | n.a. |
| HD187691 | 8.66 | 4.48 | n.a. | n.a. | 0.1483 | 0.0045 | 0.0040 | 0.1487 | 0.0059 | 0.5130 | G | 5.4F | n.a. |
| HD188512 | 4.69 | 2.04 | n.a. | n.a. | 0.1365 | 0.0033 | 0.0031 | 0.1367 | 0.0037 | 0.4449 | P | 4.1P | n.a. |
| HD190007 | 14.11 | 2.50 | n.a. | n.a. | 0.6926 | 0.0901 | 0.0842 | 0.6846 | 0.0971 | 0.4248 | P | 13.7F | n.a. |
| HD190406 | 16.35 | 6.42 | 2.62 F | n.a. | 0.1962 | 0.0110 | 0.0093 | 0.1948 | 0.0163 | 0.4770 | E | 2.6F | 15.8C |
| HD201091 | 7.19 | 36.02 | n.a. | n.a. | 0.6462 | 0.0755 | 0.0558 | 0.6389 | 0.1427 | 0.4830 | E | 7.3E | 6.95S |
| HD201092 | 11.20 | 6.17 | n.a. | n.a. | 0.9580 | 0.0946 | 0.0828 | 0.9608 | 0.1364 | 0.4409 | E | 11.7G | 13.4C |
| HD219834A | 17.28 | 10.87 | n.a. | n.a. | 0.1544 | 0.0072 | 0.0055 | 0.1536 | 0.0125 | 0.5291 | E | 21G | n.a. |
| HD219834B | 8.92 | 14.83 | n.a. | n.a. | 0.1941 | 0.0210 | 0.0155 | 0.1909 | 0.0413 | 0.5511 | E | 10.0E | 9.29S |
| HD224930 | 14.30 | 3.53 | n.a. | n.a. | 0.1866 | 0.0091 | 0.0080 | 0.1867 | 0.0127 | 0.5311 | F | 10.2P | n.a. |

**Notes.** Column (1) provides the star name, column (2) the cycle period (in years), column (3) the negative logarithm of the p-value, column (4) a secondary cycle period (in years) when present, column (5) a tertiary cycle period (in years) when present, column (6) the observed mean S-value, column (7) the dispersion thereof, column (8) the mean modeled S-value, column (9) the observed dispersion around the model cycle, column (10) the modeled peak-to-peak variations of the S-value, and column (11) the modeled phase difference between cycle maximum and minimum. Column (12) gives our cycle classification (E, G, F, and P). Column (13) provides – for comparison – the results derived by Baliunas et al. (1995) derived from the Mount Wilson data alone for the respective star, and column (14) provides the respective results derived by Oláh et al. (2016) whenever available.

## Appendix F: Cycle analysis results: Phased S-index curves

In the following figures we present three exemplary phased light curves of sample stars with "excellent," "good," "fair," and "poor" cycles according to our classification scheme. For better readability the periods used to phase the S-index data are provided on top of each individual figure; data points derived from Mount Wilson observations are color-coded in blue, data points color-coded in red are derived from TIGRE observations.

### *F.1. The "excellent" cycle stars*

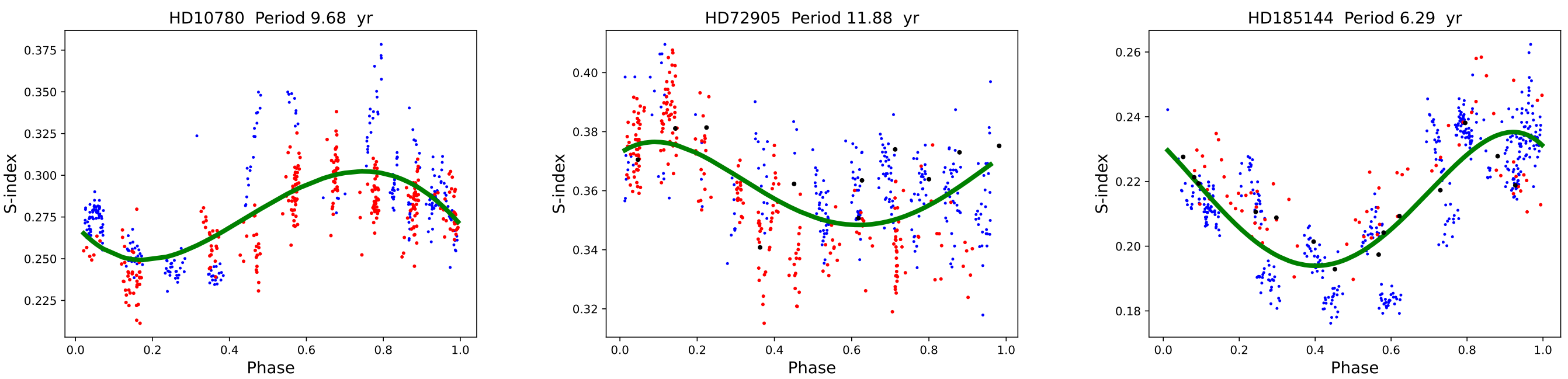

Fig. F.1: Exemplary phased SMWO-TIGRE light curves of stars with "excellent" cycles

### *F.2. The "good" cycle stars*

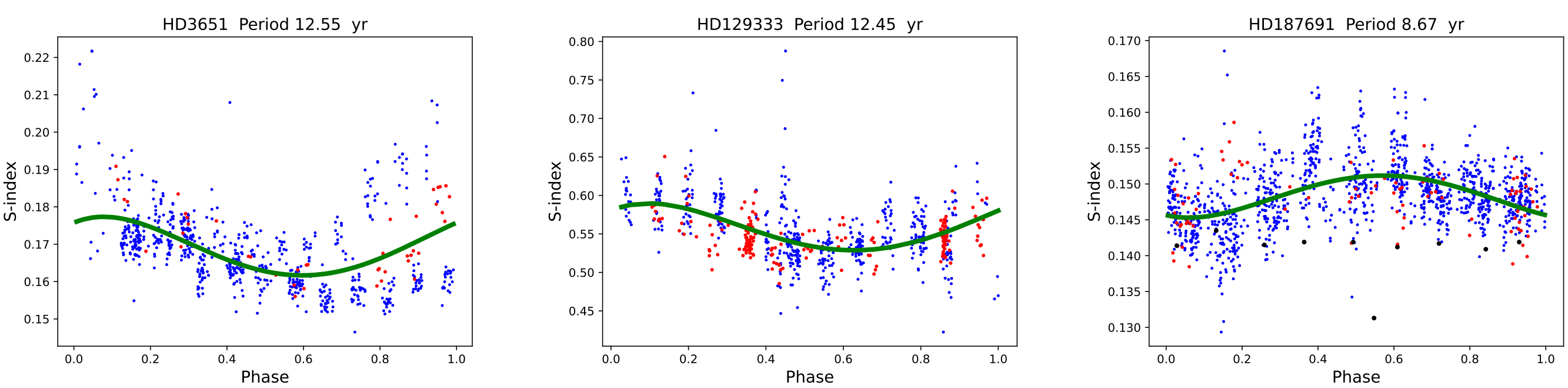

Fig. F.2: Exemplary phased SMWO-TIGRE light curves of stars with "good" cycles

### F.3. The "fair" cycle stars

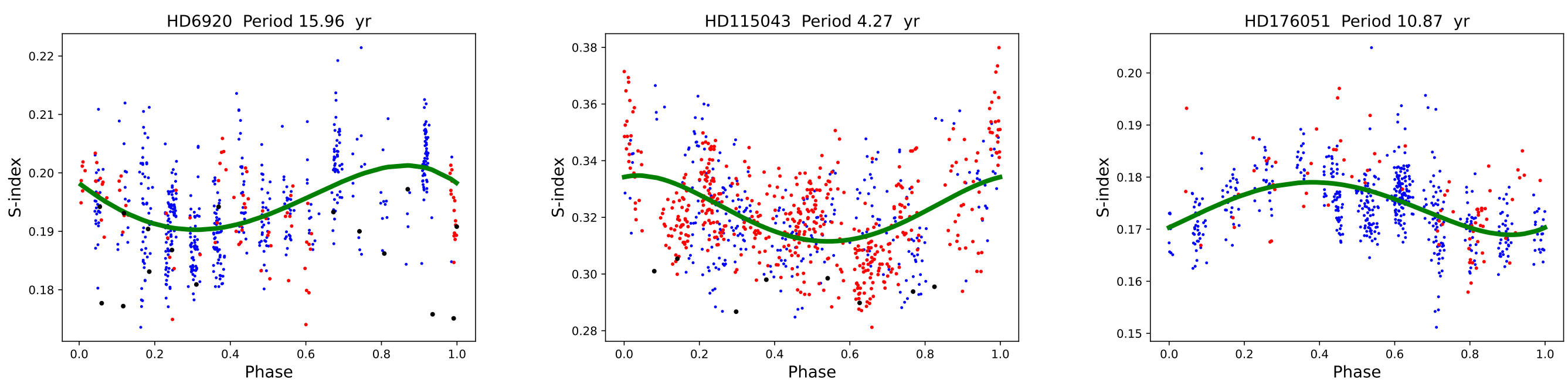

Fig. F.3: Exemplary phased SMWO-TIGRE light curves of stars with "fair" cycles.

### *F.4. The "poor" cycle stars*

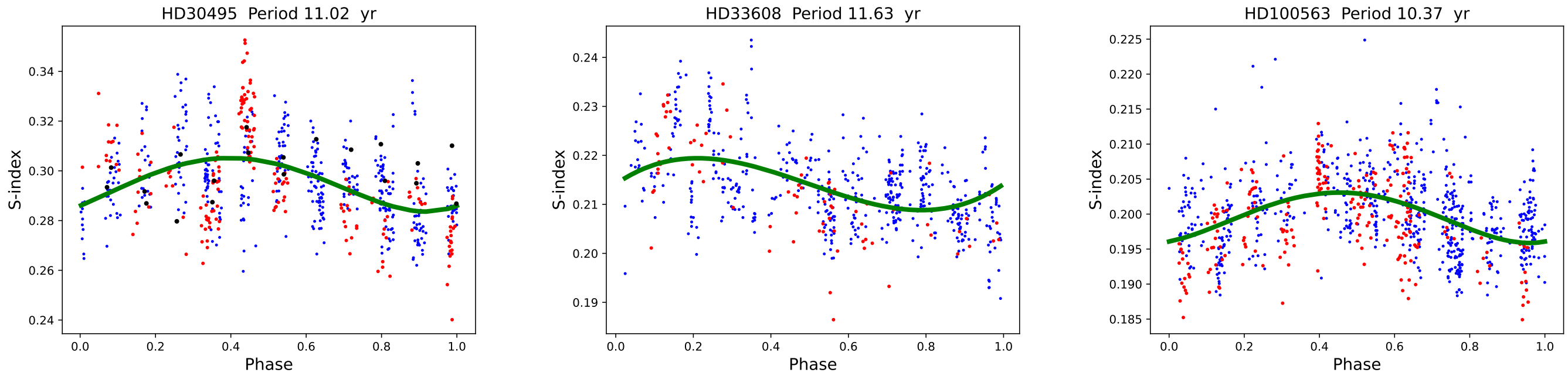

Fig. F.4: Exemplary phased SMWO-TIGRE light curves of stars with "poor" cycles.

### *F.5. Maunder minimum candidate stars*

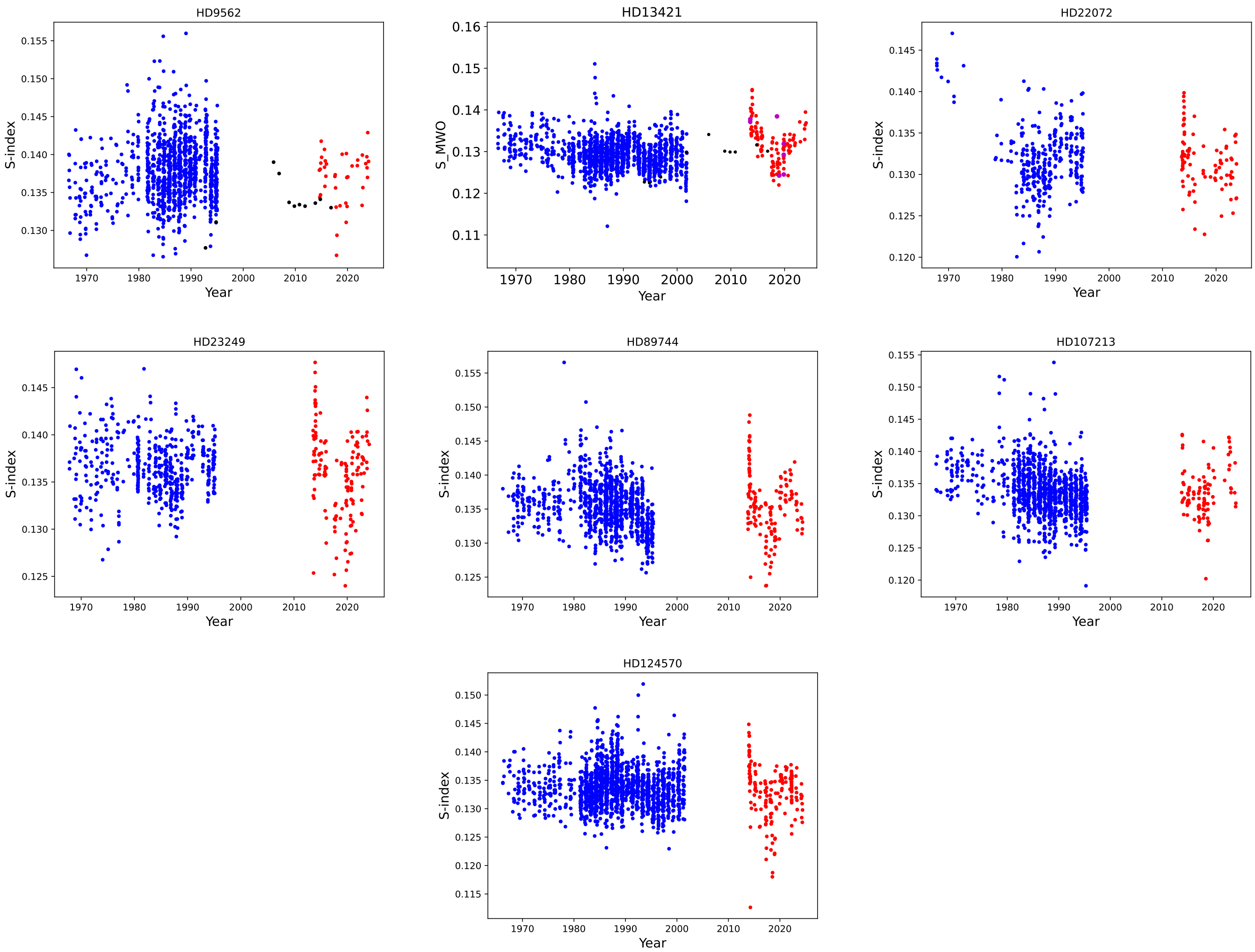

Fig. F.5: S-index light curves for the Maunder minimum candidate stars HD 9562, HD 13421, HD 22072, HD 23249, HD 89744, HD 107213, and HD 124570. Mount Wilson data points are shown with the blue data points; the TIGRE data are shown with the red data points; and the SSS seasonal medians are shown with the black data points.

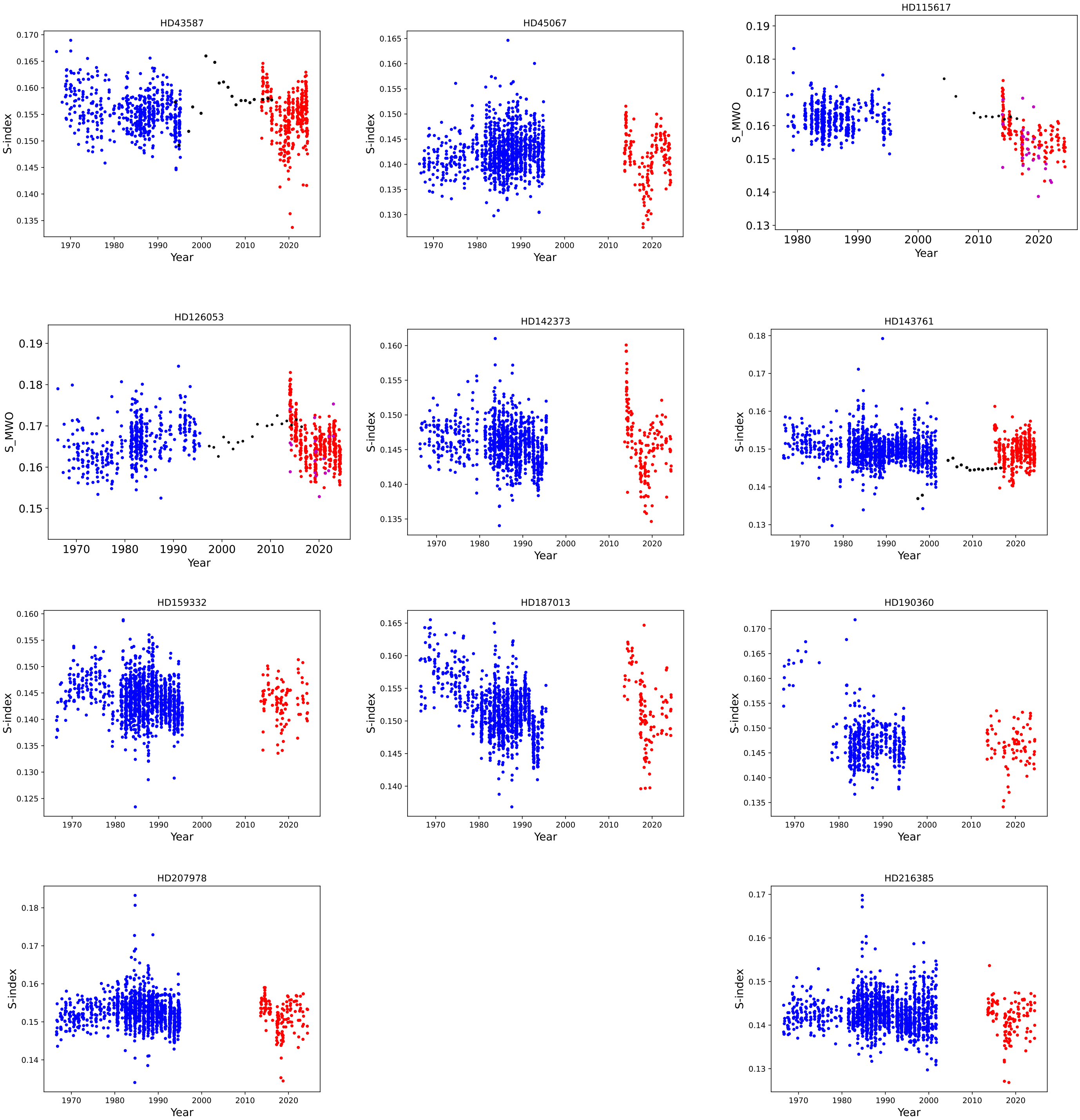

Fig. F.6: S-index light curves for the Maunder minimum candidate stars HD 43587, HD 45067, HD 115617, HD 126053, HD 142373, HD 143761, HD 159332, HD 187013, HD 190360, HD 207978, and HD 216385. Mount Wilson data points are shown with blue data points; TIGRE data are shown with red data points; and the SSS seasonal medians are shown with black data points whenever available.

In Fig. F.5 we present the joint Mount Wilson-TIGRE S-index light curves for the Maunder minimum candidate stars with S-indices below 0.14 for the stars HD 9562, HD 13421, HD 22072, HD 23249, HD 29645, HD 89744, HD 107213, HD 124570 and HD 188512 discussed in Sect. 7. In Fig. F.6 we present the joint Mount Wilson-TIGRE S-index light curves for the Maunder minimum candidate stars with S-indices between 0.14 and 0.16 for the stars HD 43587, HD 45067, HD 115617, HD 126053, HD 142373, HD 143761, HD 159332, HD 187013, HD 190360, HD 207978, HD 216385, i.e., stars with activity values below the Sun and also discussed in Sect. 7.